\documentclass[aps,prc,preprint,tightenlines,showpacs,floatfix,%
amssymb,byrevtex,nofootinbib]{revtex4}  

\usepackage{epsfig,bm,dcolumn}

\begin{document}

\preprint{hep-ph/0412363}


\title{Pentaquark $\bm{\Theta^+(1540)}$ production in
$\bm{\gamma N \to K\overline{K} N}$ Reactions}


\author{Yongseok Oh}%
\email{yoh@physast.uga.edu}

\author{K. Nakayama}%
\email{nakayama@uga.edu}

\affiliation{Department of Physics and Astronomy,
University of Georgia, Athens, Georgia 30602, U.S.A.}

\author{T.-S.H. Lee}%
\email{lee@phy.anl.gov}

\affiliation{Physics Division, Argonne National Laboratory, Argonne,
Illinois 60439, U.S.A.}

\date{\today}


\begin{abstract}

The recent developments in the search of exotic pentaquark hadrons are
briefly reviewed. We then focus on investigating 
how the exotic pentaquark $\Theta(1540)$ baryon production
can be identified in the $\gamma N \to K \overline{K} N$ reactions, focusing
on the influence of the background (non-$\Theta$ production) mechanisms.
By imposing the SU(3) symmetry and using various quark model predictions, we
are able to fix the coupling constants for evaluating the kaon
backgrounds, the $K\overline{K}$ production through the intermediate
vector meson and tensor meson photoproduction, and the mechanisms involving
intermediate $\Lambda(1116)$, $\Lambda(1405)$, $\Lambda(1520)$,
$\Sigma (1193)$, $\Sigma(1385)$, and $\Delta (1232)$ states.
The vector meson photoproduction part is calculated from a phenomenological
model which describes well the experimental data at low energies. 
We point out that the neutral tensor meson production can not be due to
$\pi^0$-exchange as done by Dzierba {\it et al.\/} [Phys. Rev. D {\bf
69}, 051901 (2004)] because of $C$ parity. 
The neutral tensor meson production is estimated by considering the vector
meson exchange and found to be  too weak to generate any peak at the
position near $\Theta(1540)$.
For $\Theta(1540)$ production, we assume that it is an isoscalar and hence
can only be produced in $\gamma n \rightarrow K^+K^-n$ and
$\gamma p \rightarrow K^0\overline{K}^0p$ reactions, but not in
$\gamma p \rightarrow K^+K^-p$ and $\gamma n \rightarrow K^0\overline{K}^0 n$.
The total cross section data of $\gamma p \rightarrow K^+K^- p$ is thus used
to fix the form factors which regularize the background amplitudes so that
the signal of $\Theta(1540)$ in $\gamma n \rightarrow K^+K^-n$ and 
$\gamma p \rightarrow K^0\overline{K}^0p$ cross sections can be predicted.
We find that the predicted $K^+K^-$ and $K^+ n$ invariant mass distributions
of the $\gamma n \to K^+ K^- n$ reaction can qualitatively reproduce the
shapes of the JLab data.
However, the predicted $\Theta(1540)$ peak can not be identified unambiguously
with the data.
High statistics experiments are needed to resolve the problem.
We also find that an even-parity $\Theta$ is more likely to be detected, while
it will be difficult to identify an odd-parity $\Theta$, even if it exists,
from the background continuum, if its coupling constants are small as in
the present quark model predictions.

\end{abstract}


\maketitle

\section{Introduction}

The recent interests in pentaquark baryons was initiated by the discovery
of $\Theta^+(1540)$ by the LEPS Collaboration at SPring-8~\cite{LEPS03}
and the subsequent
experiments~\cite{DIANA03,CLAS03-b,SAPHIR03,CLAS03-c,ADK03,CLAS03-d,%
AMW03,HERMES-04,SVD-04,SVD-05,AER04,COSY-04,ZEUS04d}.
Candidates for exotic pentaquark states $\Xi(1862)$ and $\Theta_c(3099)$
were also observed~\cite{NA49-03,H1-04}.
However, the signals for those exotic states could not be found in several
recent experiments~\cite{SPHINX04,PHENIX-04,CDF04b,HyperCP-04,BABAR04c,%
BABAR04b,HERA-B-04b,BES04c,FOCUS-04,CDF04,ALEPH-04,Belle04,CLAS05,WA89-05}.
In addition, more reports have been given on possible candidates for various
crypto-exotic
baryons~\cite{Lands99d,SPHINX02,BES04b,GRAAL04,ZACC04,Tati04,STAR04}.
The existence of pentaquark baryons is thus not conclusive at the present
time.%
\footnote{The most recent experimental report from the STAR
Collaboration~\cite{STAR05} would suggest the existence of a narrow
$\Theta$ with isospin $I\neq 0$ at a mass $1528 \pm 2 \pm 5$ MeV.}
In spite of this situation, it seems worthwhile at this stage to collect
together in one place the various references to the literature with a
brief review of major experimental and theoretical works in this rapidly
expanding area of investigation.

\begin{figure}[t]
\centering
\epsfig{file=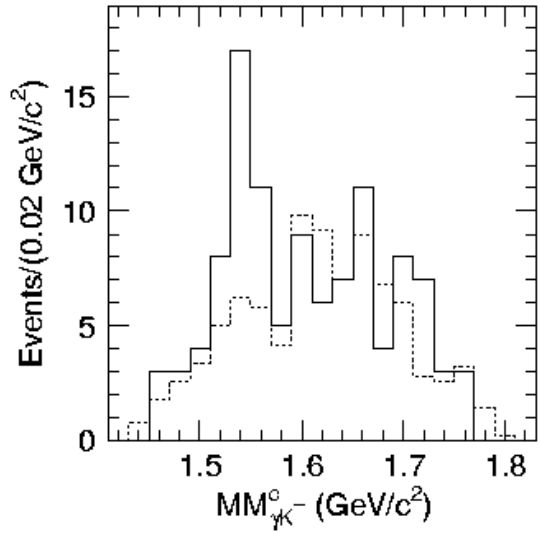, width=0.5\hsize}
\caption{Data of missing mass spectrum of ${}^{12}C(\gamma K^+ K^-)$ reaction
from the experiment at LEPS \cite{LEPS03}.}
\label{fig:LEPS}
\end{figure}

\begin{figure}[t]
\centering
\epsfig{file=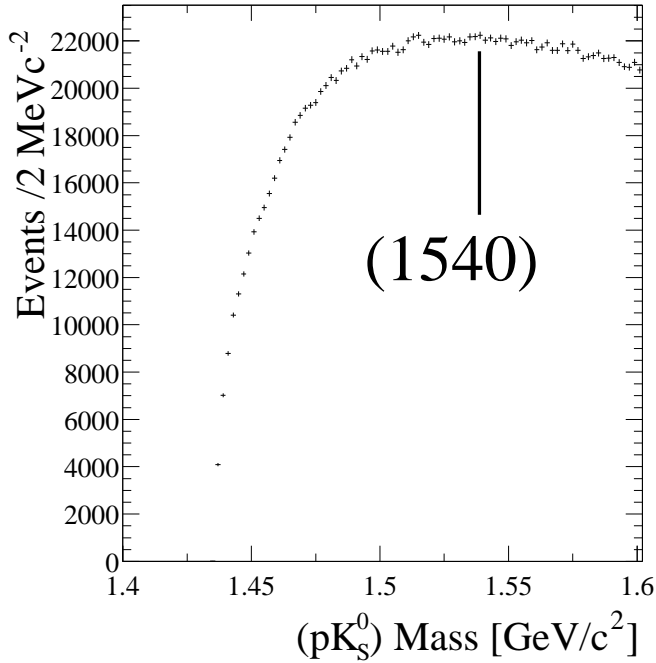, width=0.6\hsize}
\caption{Data from 
 $e^+e^- \rightarrow p K^0 X$ experiment at BaBar \cite{BABAR04b}.}
\label{fig:BaBar}
\end{figure}

The results from recent experiments are summarized in Tables
\ref{tab:positive} and \ref{tab:negative}. 
We see that the positive (Table~\ref{tab:positive}) and negative
(Table~\ref{tab:negative})
reports on the existence of $\Theta^+(1540)$ are almost equally divided.%
\footnote{For the negative reports on the existence of $\Xi(1860)$ and
$\Theta_c(3099)$, see, e.g., Refs.~\cite{FW04,WA89-04,Pocho04} and the
references in Table~\ref{tab:negative}.}
In Fig.~\ref{fig:LEPS}, the peak of the spectrum at the missing mass 
$\sim 1.54$ GeV was identified by the LEPS group with the excitation of
 $\Theta^+(1540)$ in the $\gamma n \to K^+ K^- n$ process taking place
in  the $^{12}$C target.
On the other hand,  the spectrum obtained from the $e^+ e^- \to p K^0 X$
experiment at BaBar \cite{BABAR04b} does not show any resonance peak at
the predicted position of $\Theta^+(1540)$ as seen in Fig.~\ref{fig:BaBar}.

\begin{table}[b]
\centering
\begin{tabular}{cc|c|c} \hline \hline
\multicolumn{2}{c}{Expt.} &  Reaction &    Reference \\ \hline
LEPS & (2003) &   $\gamma + {}^{12}\mbox{C}$  & \cite{LEPS03}   \\
DIANA & (2003) &  $K^+ + \mbox{Xe}$  & \cite{DIANA03} \\
CLAS & (2003) &  $\gamma + d$ & \cite{CLAS03-b} \\
SAPHIR & (2003) & $\gamma + p$ & \cite{SAPHIR03} \\
BBCN & (2003) & $\nu_\mu (\overline{\nu}_\mu) + A$ & \cite{ADK03} \\
CLAS & (2003) & $\gamma + p$ & \cite{CLAS03-d} \\
HERMES & (2003) & $e+ d$ & \cite{HERMES-04} \\
SVD & (2004) &  $p+A$ & \cite{SVD-04} \\
COSY-TOF & (2004) & $p + p$ & \cite{COSY-04} \\
ZEUS & (2004) & $e^\pm + p$ & \cite{ZEUS04d} \\
JINR & (2004) & $p + A$ & \cite{AER04} \\
SVD & (2005) &  $p+A$ & \cite{SVD-05} \\
\hline\hline
\end{tabular}
\caption{Positive reports on the evidence of $\Theta^+(1540)$ pentaquark 
state.}
\label{tab:positive}
\end{table}
                                                                                
\begin{table}[t]
\centering
\begin{tabular}{cc|c|c} \hline \hline
\multicolumn{2}{c}{Expt.} &  Reaction &    Reference \\ \hline
BES & (2004) & $e^+ + e^- $ & \cite{BES04c} \\
PHENIX & (2004) & $A + A $ & \cite{PHENIX-04} \\
ALEPH & (2004) & $e^+ + e^- $ & \cite{ALEPH-04} \\
SPHINX & (2004) & $p + A$ & \cite{SPHINX04} \\
BaBar  & (2004)& $e^+ + e^- $ & \cite{BABAR04b} \\
CDF & (2004) & $p + \bar{p}$ & \cite{CDF04b} \\
HERA-B & (2004) & $p+ A$ & \cite{HERA-B-04b}\\
HyperCP & (2004) & $p +A$ & \cite{HyperCP-04} \\
Belle & (2004) & $e^+ + e^- $ & \cite{Belle04} \\
FOCUS & (2004) & $\gamma + A$ & \cite{FOCUS-04} \\
CLAS & (2005) & $\gamma + p$ & \cite{CLAS05} \\
WA89 & (2005) & $\Sigma^- + A$ & \cite{WA89-05} \\
\hline\hline
\end{tabular}
\caption{Negative reports on the evidence of $\Theta^+(1540)$ pentaquark
state.}
\label{tab:negative}
\end{table}

It is useful to first briefly review all of the theoretical works on 
pentaquark baryons which could be pure exotic or crypto-exotic.
The pure exotic states can easily be identified by their unique quantum
numbers, but the crypto-exotic states are hard to be identified as their
quantum numbers can also be generated by three-quark states.
Therefore, it is crucial to have careful analyses for their decay
channels and other properties.
Historically, there have been many efforts to find pentaquark states with
the development of quark models, which, however, failed to observe
$\Theta(1540)$.
The efforts to search for pentaquark baryons until 1980's were summarized in
Refs.~\cite{Golo71,PDG86}. (See also Ref.~\cite{GM99}.)%
\footnote{
In the literature we could find several resonances that were claimed to be
crypto-exotic states.
For example, $X(1340)$, $X(1450)$, and $X(1640)$ were reported by
Refs.~\cite{HKKK79,FHHK80} and $X(3520)$ by Ref.~\cite{KMCM91}.
$X(1390)$, $X(1480)$, and $X(1620)$ that have isospin $I \ge 5/2$ were
observed by Ref.~\cite{ABGG90}, and Ref.~\cite{BCGMP79} reported
$\Sigma(3170)$.
Most of them were found to have narrow widths, but their existence was not
confirmed and questioned by later experiments \cite{AGDD91,ACDD85}.
SPHINX Collaboration has reported the existence of $X(2000)$, $X(2050)$,
and $X(2400)$ that are expected to have the quark content of $uuds\bar{s}$
\cite{Lands99d,SPHINX02}, whose existence should be carefully re-examined
by other experiments.}
Early theoretical works on exotic baryons can be found, e.g., in
Refs.~\cite{Jaf77,Strot79,HS78,DHS79,MG86,FPG92}.

Rigorous theoretical studies were then performed for heavy quark sector,
i.e., pentaquark baryons with one anti-charmed quark or anti-bottom
quark.
In the pioneering work of Lipkin \cite{Lip87} and Grenoble group
\cite{GSR87}, the anti-charmed pentaquark with one strange quark was
shown to have the same binding energy as the $H$ dibaryon in the heavy
quark mass limit and in the SU(3) limit.
Then it has been studied in more sophisticated quark models
\cite{FGRS89,ZR94,GRSP97,Stan98}, which improved the simple prediction
of Refs.~\cite{Lip87,GSR87} and some of them predicted no bound state.
The heavy pentaquark systems are also investigated using the Skyrme model,
which gives different results compared with the quark model.
In this approach, the bound-state model of Callan and
Klebanov~\cite{CK85} was applied to study the heavy pentaquark system
after it was shown that the model can be successfully applied to the
normal heavy-quark baryons~\cite{RRS90,OMRS91}.
In Ref.~\cite{RS93}, Riska and Scoccola used the Skyrme Lagrangian with
symmetry breaking terms to investigate the heavy pentaquark system and found
that some of the non-strange heavy pentaquarks can be deeply bound and
therefore stable against the decays by strong interaction.
This is quite a remarkable result compared with the quark model where a
nonstrange pentaquark baryon has no sufficient symmetry to be stable
via the hyperfine interactions.
However, this model does not satisfy the heavy quark symmetry at the infinite
heavy quark mass limit \cite{VS87,IW89} by integrating out the heavy vector
meson field.
It was shown that including the heavy vector meson fields explicitly is
essential for satisfying heavy quark symmetry \cite{JMW93a}, and the model
was successfully applied to heavy quark baryons~%
\cite{GLM93,MOPR95,OPM94,SS95,HSSW97,OPM94b,OP95,OP96,OP97}.
This model was then used to study the heavy pentaquarks in
Refs.~\cite{OPM94a,OPM94b}, which gives stable nonstrange pentaquark
baryons, although the binding energy and the mass formulas are quite
different from those of Ref.~\cite{RS93}.
The finite mass corrections and the soliton-recoil effects are discussed in
Refs.~\cite{OP95,OP97}.
The extension to strange heavy pentaquarks can be found in Ref.~\cite{Chow95}.
Following the first experimental search for heavy pentaquarks
\cite{E791-98,E791-99}, the observation of $\Theta^+(1540)$ and
$\Theta_c(3099)$ has brought new interests in the heavy pentaquarks~%
\cite{Cheung03,CCH04,HL04,BKM04,SWW04,WM04c,KLO04}.

In the light quark sector, pentaquark states were anticipated already in the
Skyrme model \cite{Chem85,Man84,Pras87,Weig98}.
The first detailed study on antidecuplet was made by Diakonov {\it
et al.\/} \cite{DPP97,PSTCG00}, which predicted a very narrow $\Theta^+$
with a mass around 1530 MeV by identifying $N(1710)$ as the nucleon
analogue of the antidecuplet.
After the discovery of $\Theta^+(1540)$ there have been lots of
theoretical models and ideas to explain the structure of pentaquark
baryons and to search for the other pentaquark states.
The subsequent theoretical studies include the soliton models
\cite{WK03,Pras03,IKOR03,JM03,PRM04,PG05,Cohen03,CCDL05,WW05}, QCD sum rules
\cite{Zhu03,MNNRL03,SDO03,Eidem04}, large $N_c$ QCD
\cite{CL04,JM04a,JM04b,Man04}, and
lattice calculation \cite{CFKK03,Sasaki03,MLAB04,CH04a}, etc.

As the quark models have provided a cornerstone for hadron physics, it
is legitimate to start with the quark models and study the structure
of pentaquark baryons.
In Ref.~\cite{KL03a}, Karliner and Lipkin suggested a triquark-diquark model,
where, for example, $\Theta^+$ is a system of $(ud)$-$(ud\bar{s})$.
In Ref.~\cite{JW03}, Jaffe and Wilczek advocated a
diquark-diquark-antiquark model so that $\Theta^+$ is $(ud)$-$(ud)$-$\bar{s}$.
In this model, they also considered the mixing of the pentaquark
antidecuplet with the pentaquark octet, which makes it different from the
SU(3) soliton models where the octet describes the normal (three-quark)
baryon octet.
Assuming that the nucleon and $\Sigma$ analogues are in the ideal mixing
of the octet and antidecuplet, the nucleon analogue is then identified as
the Roper resonance $N(1440)$.
In Ref.~\cite{OKL03b}, however, it was pointed out that the 
$N(1710)$ should be excluded as a pure antidecuplet state.
This is because, within SU(3) symmetry, antidecuplet does not couple to
decuplet and meson octet, whereas $N(1710)$ has a large branching ratio into
$\pi\Delta$ channel.
Therefore, mixing with other multiplets is required if one wants to identify
$N(1710)$ as a pentaquark crypto-exotic state.
However, recent study for the ideal mixing between antidecuplet and octet
states shows that the ideally mixed state still has vanishing coupling with
the $\pi\Delta$ channel \cite{CD04,LKO04}, which excludes $N(1440)$ as a
pentaquark state.
This shows the importance of reaction/decay studies in identifying
especially crypto-exotic pentaquark states.
More discussions on the quark model predictions based on the diquark
picture can be found, e.g., in Refs.~\cite{OKL03b,DP03b}.
Predictions on the antidecuplet spectrum in various quark models can be
found, e.g., in Refs.~\cite{BGS03,SR03,CCKN03a,CCKN04b,GK03b,KLV04}

In quark model, pentaquark baryons form six multiplets, $\bm{1}$,
$\bm{8}$, $\bm{10}$, $\overline{\bm{10}}$, $\bm{27}$ and $\bm{35}$.
The other type resonances are thus expected together with antidecuplet,
particularly the isovector $\Theta$ belonging to $\bm{27}$-plet and
isotensor $\Theta$ as a member of $\bm{35}$-plet.
The interest in this direction has been growing
\cite{BGS03,Pras04,WM04a,EKP04,DS04} and
it is important to know the interactions and decay channels to
search for the other pentaquark baryons if they exist.

Furthermore, understanding the $\Theta^+$ properties such as spin-parity
requires careful analyses of production mechanisms%
\footnote{
See, however, Refs.~\cite{RT04a,NL04,THH03,HBEK03,RT04b,U04,HHNM04}, 
where efforts to determine these properties, in particular the parity of the
$\Theta^+$, in a model-independent way in photo- and hadro-production
reactions have been reported.} 
including $\gamma N \to \bar{K} \Theta$
\cite{LK03a,LK03b,NHK03,OKL03a,ZA03,YCJ03},
$\gamma N \to \bar{K}^* \Theta$ \cite{OKL03d},
$\gamma N \to K \pi \Theta$ \cite{LKK03}
$\gamma N \to K^+ K^- N$
\cite{NT03,DKST03,Roberts04,Roberts04b,TEHN04,SHKMT05},
$NN \to Y\Theta$ \cite{OKL03c,THH03,HBEK03,HHNM04}, and
$K N \to K \pi N$ \cite{HHO03}.
Most model predictions for those production processes, however, do not
consider the intermediate pentaquark baryons in its production mechanisms
as the unknown inputs like the electromagnetic and strong couplings
of pentaquark baryons are required.
Therefore, knowing the interaction Lagrangian of pentaquark baryons are
necessary for understanding the production mechanisms.

The physical pentaquark states would be mixtures of various multiplets
as in the chiral soliton model \cite{EKP04}.
Such a representation mixing is induced by SU(3) symmetry breaking
and it can be studied in quark potential models.
Therefore, it is desirable to obtain the full set of pentaquark wave
functions in quark model for further investigation.
There are several works in this direction and the flavor wave functions
of antidecuplet has been obtained in
Refs.~\cite{CCKN03a,CCKN04b,CCKN03c,CD04}.
(See also Refs.~\cite{JM04a,JM04b,Man04} for the relation between the wave
functions of
pentaquark baryons in quark model and Skyrme model in the large $N_c$
limit.)
The SU(3) symmetric interactions for antidecuplet have been studied in
Refs.~\cite{OKL03b,LKO04,CD04}, which motivated the development of a
chiral Lagrangian for antidecuplet \cite{KLLP03,LZHD04}.
In a recent paper \cite{OK04}  the SU(3) quark model has been extended
to obtain the flavor wave functions of all pentaquark states including
singlet, octet, decuplet, antidecuplet, $\bm{27}$-plet and
$\bm{35}$-plet.
Thus  the SU(3) symmetric Lagrangian describing the 
pentaquark--three-quark and pentaquark-pentaquark interactions with
meson octet has been obtained.

{}From the above brief descriptions of theoretical works on pentaquark
baryons, it is clear that the verification of the existence of
$\Theta^+(1540)$ is a very important step in the development of hadron physics.
To make progress, both experimental and theoretical efforts are needed.
Experimentally, several high statistics experiments have been
planned~\cite{CLAS04c,Hicks04,VDN04,KS04,Oha04} and some data have
been obtained and are under analyses.
On the theoretical side, it is necessary to understand the reaction
mechanisms of the considered reactions and to investigate how
the pentaquark states production can be identified from the
experimental observables.
In particular, one must explore whether the resonance-like peaks near
$W \sim 1540$ MeV can be resulted from the background
(non-$\Theta$ production) mechanisms, as emphasized by Dzierba
{\it et al.\/}~\cite{DKST03}.
In this paper, we report on the progress we have made in this direction
concerning the $\gamma N \rightarrow K\overline{K} N$ reactions.

The $\gamma N \rightarrow K\overline{K} N$ reaction has been used to observe
the production of $\Theta(1540)$ through its decay into
$KN$~\cite{LEPS03,CLAS03-b,SAPHIR03,HERMES-04,ZEUS04d,HERA-B-04b}.%
\footnote{Old experiments to find the pentaquark states using this
reaction can be found, e.g., in Ref.~\cite{TGHL67}.}
We will take the effective Lagrangian approach to assume that the amplitudes
of this reaction can be computed from the tree-diagrams.
In addition to the $\Theta$ production amplitude, we compute all possible
background (non-$\Theta$ production) amplitudes whose parameters can be
fixed by the SU(3) symmetry or taken from various quark model predictions.
Undoubtedly, there are some tree-diagrams which are kinematically allowed,
but can not be evaluated because of the lack of information about the relevant
coupling constants.
So our effort represents just a step toward developing a model for more
realistic production amplitudes.
But it should be sufficient for exploring the question concerning how the
pentaquark $\Theta(1540)$ can be identified from the observables of
$\gamma N \rightarrow K\overline{K} N$ reaction.

We classify the possible tree diagrams into four types: the (tree) $t$-channel
Drell diagrams (Fig.~\ref{fig:Drell}),
vector meson and tensor meson
production background (Fig.~\ref{fig:VTM}), hyperon background
(Fig.~\ref{fig:hyp}), and the $\Theta(1540)$ production amplitude
(Fig.~\ref{fig:theta}).%
\footnote{
The full amplitudes of the diagrams of Fig.~\ref{fig:Drell} together with
the hyperon and $\Theta$ production diagrams constitute the so-called Drell
diagrams~\cite{Soeding66}, where
the incoming photon is converted into a $K\overline{K}$ pair or a
$K\overline{K}^*$ ($\overline{K}K^*$) pair and then a virtual $K$
(or $K^*$) diffractively scatters on the nucleon
\cite{Drell60,Drell61}.}
Our objective is to explore the 
interplay between these tree diagrams in determining  
the following reactions:
\begin{eqnarray}
& \gamma p \to K^+ K^- p, \qquad&  \gamma n \to K^+ K^- n,
\nonumber \\
& \gamma p \to K^0 \overline{K}^0 p, \qquad&  \gamma n \to K^0
\overline{K}^0 n.
\label{eq:channel}
\end{eqnarray}
The parameters as well as the form factors for evaluating these diagrams
will be explained in detail in Section~II and in Appendices.

There exist several investigations of the reactions listed in
Eq.~(\ref{eq:channel}) in conjunction with the observation of
$\Theta(1540)$~\cite{NT03,DKST03,Roberts04,TEHN04}.
(See also Refs.~\cite{SS04b,DMS04}.)
The background tree-diagrams considered by these works are summarized in
Table~\ref{tab:models}.
In the very first investigation of $\gamma n \rightarrow K^+K^- n$ by
Nakayama and Tsushima (NT)~\cite{NT03}, the kaon background, vector meson
background, and hyperon background are considered.
But the tensor meson background was not included.
Their hyperon background includes the $\Sigma(1193)$ and $\Sigma(1660)$
states, while the $\Lambda$ baryon resonances do not come into play in the
considered $\gamma n \to K^+ K^- n$ reaction because of the isospin
selection rule.

The investigation of the $\gamma n \to K^+ K^- n$ reaction by Dzierba
{\it et al.\/} (DKSTS)~\cite{DKST03} was motivated by an old
experiment~\cite{ABBC69} on $\pi^- p \to K^- X$ reaction which was aimed at
searching for the $\Theta$ baryon(s).
With 8 GeV pion beams, this experiment found two peaks at $W = 1590$ and
$1950$ MeV in the $KN$ channel.
However, these peak positions moved to 1500 and 1800 MeV as the pion beam
energy was lowered to 6 GeV.
They hence concluded that the most natural explanation was to ascribe the
peaks to the background production mechanisms, especially to the
higher-spin meson production.
Motivated by this analysis, Dzierba {\it et al.\/} claimed the
possibility that the peak at 1540 MeV in $\gamma n \to K^+ K^- n$ may
come from the tensor meson photoproduction background.
However, other production mechanisms such as $t$-channel Drell diagrams and
hyperon backgrounds were not included in their investigation, as indicated in
Table~\ref{tab:models}.
We will discuss this work in more detail later.

In Ref.~\cite{Roberts04}, Roberts studied the contributions of $\Theta$ to
the invariant mass distributions of $\gamma N \to K\overline{K}N$
reactions listed in Eq.~(\ref{eq:channel}). 
In addition to including many $\Lambda$ and $\Sigma$ baryons, he included
vector meson production, but neglected the tensor meson production and
the photo-transitions in the intermediate baryon states.
He also explored how the $\gamma N \to K\overline{K}N$ observables can be
used to distinguish the spin-parity of $\Theta(1540)$ by considering 
$J^P = \frac12^\pm$ or $\frac32^\pm$ for $\Theta(1540)$.
However, a common form factor was used for simplicity and as a result
the amplitudes are not constrained by experimental data, which should be
improved for more realistic models.

Titov {\it et al.\/} (TEHN)~\cite{TEHN04} investigated the effects of the
scalar meson, vector meson, and tensor meson production background on the
$\gamma N \to K \overline{K} N$ reactions.
They found that the $\phi$ meson photoproduction due to the Pomeron-exchange
is the major background mechanism.
This result was obtained from using a phenomenological model which includes
the meson exchanges and Pomeron exchange and can describe well the $\phi$
photoproduction data. 
The main emphasis in Ref.~\cite{TEHN04} was to explore the sensitivity and/or
insensitivity of various spin asymmetries to the parity of $\Theta$ by
improving on the work of Ref.~\cite{NT03}.
However, the $t$-channel Drell diagrams and hyperon background were not
included in this investigation.
In addition, this work focused only on a special kinematic region, i.e.,
at the resonance point, so the invariant mass distribution, which is required
to verify the existence of the $\Theta$ state, could not be discussed,

\begin{table}
\centering
\begin{tabular}{c|ccccc} \hline\hline
Models & $t$-channel Drell & Vector meson & Tensor meson
& Hyperon &  \\
      & diagrams & background & background & background &  \\  \hline
NT~\cite{NT03} & $\surd$ & $\surd$\, & & $\surd$  &  \\
DKSTS~\cite{DKST03} & & $\surd$ & $\surd$ & & \\
Roberts~\cite{Roberts04} & $\surd$ & $\surd$\ &  & $\surd$ &  \\
TEHN~\cite{TEHN04} &  & $\surd$ & $\surd$ & &  \\ 
This work &$\surd$  & $\surd$ & $\surd$ &$\surd$ &  \\
\hline\hline
\end{tabular}

\caption{Comparison of the models for $\gamma N \to K\overline{K}N$
as a background for the $\Theta(1540)$ production in
the literature. See the text for the details.}
\label{tab:models}
\end{table}

As indicated in the last row of Table~\ref{tab:models}, we consider in
this work all four classes of the background amplitudes.
Therefore, the present work represents an extension of the work initiated by
Nakayama and Tsushima~\cite{NT03} toward developing a more realistic model
of the $\gamma N \to K\overline{K} N$ reactions in order to extract relevant
information concerning the pentaquark $\Theta$.
The main challenge here is to find the coupling constants which are needed to
evaluate all possible tree diagrams. Undoubtedly, no progress can be made
unless some truncations and approximations are taken.
We impose the SU(3) symmetry and only keep the tree diagrams whose coupling
constants can be either determined from using the information given by the
Particle Data Group (PDG) or taken from various quark model predictions.
Thus, we only consider the rather well-known hyperons, namely,
$\Lambda(1116)$, $\Lambda(1405)$, $\Lambda(1520)$, $\Sigma(1193)$, and
$\Sigma(1385)$ to evaluate the hyperon background (Fig.~\ref{fig:hyp}).
The photo-transitions among those hyperons and the $\Delta \to N\gamma$ 
transition are also included.
For the background amplitudes with vector meson and tensor meson
production (Fig.~\ref{fig:VTM}), the $\gamma N \to (V,T) N$ amplitudes are
generated from the available phenomenological models which are constrained by
the total cross section data of photoproduction of vector mesons
($\rho$, $\omega$, $\phi$) and tensor meson [$a_2^+(1320)$].
We also improve the model for neutral tensor meson photoproduction of
Ref.~\cite{TEHN04} by including all possible vector meson exchanges.

Another feature of our approach is regularizing each vertex in the considered
tree diagrams by a form factor which depends on the mass and squared momentum
of the exchanged particles.
Motivated by the methods of Refs.~\cite{Ohta89,Habe97,HBMF98a,DW01a},
the current conservation is recovered by introducing contact diagrams.
The details can be found in Appendix~B.
The cutoff parameters of the form factors are determined by the available
data of the $\gamma p \to K^+K^- p$ reaction, which are then used
to compute various observables.
Thus our procedure in introducing the form factors is different from all
the models listed in Table~\ref{tab:models}, such as no form factor in
the work of Ref.~\cite{NT03} and the use of the same form factor for all
diagrams in the approach of Ref.~\cite{Roberts04}.

We are also motivated by the question concerning the quantum numbers of
$\Theta(1540)$.
In Ref.~\cite{SAPHIR03}, SAPHIR Collaboration claims non-existence of
$\Theta(1540)$ in $K^+p$ channel while they could confirm the peak of
$\Theta(1540)$ in $K^+n$ channel.
This leads to the conclusion that the observed $\Theta(1540)$ is isosinglet
and it belongs to baryon antidecuplet.%
\footnote{Flavor SU(3) symmetry allows three kinds of $\Theta$ baryons
in quark models; isosinglet in antidecuplet, isovector in $\bm{27}$-plet,
and isotensor in $\bm{35}$-plet \cite{OK04}.}
However, the spin-parity quantum numbers of $\Theta$ are still under
debate.
Theoretically, uncorrelated quark models \cite{CCKN03a}, QCD sum rules
\cite{Zhu03,SDO03}, lattice QCD \cite{CFKK03,Sasaki03,MLAB04} favors
odd-parity of $\Theta(1540)$, while even-parity is predicted by correlated
quark models \cite{CCKN04b,JW03,KL03a} and soliton models \cite{DPP97}.
In addition, there is a debate on the parity of $\Theta$ in lattice
calculation \cite{CH04a} and the parity-flip was claimed in QCD sum
rules \cite{KLO04} with heavier anti-quark.
(See also Refs.~\cite{Oka04b,LKK04,RN04,LKV04} for the status of QCD sum rules
calculation.)
However, most models identify the $\Theta(1540)$ as a member of the
antidecuplet with spin-1/2.
We therefore will use this assumption in this work.
To be more specific, we follow the observations of Ref.~\cite{SAPHIR03} to
assume that $\Theta$ is of isosinglet and hence can only be produced via
the mechanisms of Fig.~\ref{fig:theta} in $\gamma n\to K^+K^- n$ and
$\gamma p \to K^0\overline{K}^0 p$ of Eq.~(\ref{eq:channel}).
The other two processes in Eq.~(\ref{eq:channel}) will also be considered
for providing information to constrain the background
(non-$\Theta$ production) amplitude by using the available data.
We will  also predict how the observables of $\gamma n\to K^+K^- n$
depend on the parity of $\Theta(1540)$, as done in
Refs.~\cite{LK03a,LK03b,NHK03,OKL03a,OKL03c,%
ZA03,NL04,YCJ03,OKL03d,THH03,HBEK03,HHO03} for the other reactions.

This paper is organized as follows.
In Section~II, we present our model for $K\overline{K}$ pair photoproduction
reactions listed in Eq.~(\ref{eq:channel}).
The form of the employed effective Lagrangians is given explicitly and
the determination of their coupling constants is also discussed in detail.
We present our results on the total and differential cross sections in
Section~III.
A comment on the spin asymmetries is also made.
A summary is given in Sect.~IV and the details for the tensor meson
properties, effective Lagrangians with the couplings, and form factors
are given in Appendixes.

\section{$\bm{K\overline{K}}$ pair photoproduction from the nucleon}

As mentioned in Section~I, we will investigate all four reactions listed in 
Eq.~(\ref{eq:channel}) by considering the tree-diagrams illustrated in
Fig.~\ref{fig:Drell} for the (tree) $t$-channel Drell mechanisms,
Fig.~\ref{fig:VTM} for the
production through vector meson and tensor meson photoproduction,
Fig.~\ref{fig:hyp} for the hyperon background, and Fig.~\ref{fig:theta} for
the $\Theta$ production.
The effective Lagrangians needed for calculating each class of these tree
diagrams will be given explicitly in the following subsections and
Appendixes. 
The determinations of the relevant coupling constants will be discussed
in detail.

\subsection{$t$-channel Drell-type diagrams}

\begin{figure}[t]
\centering
\epsfig{file=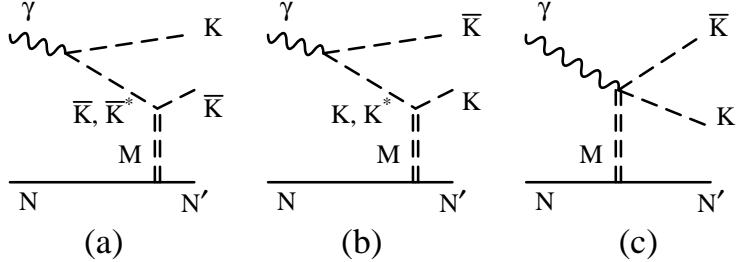, width=0.6\hsize}
\caption{Drell-type diagrams for $\gamma N \to K\overline{K} N$. Here
$M$ stands for a vector meson ($V=\rho,\omega$) or a pseudoscalar
meson ($\varphi = \pi,\eta$).}
\label{fig:Drell}
\end{figure}

For the $t$-channel Drell-type diagrams (Fig.~\ref{fig:Drell}), the incoming
photon is
converted into a $K\overline{K}$ pair or a $K\overline{K}^*$
($\overline{K}K^*$) pair and then a virtual $K$ (or $K^*$) diffractively
scatters on the nucleon \cite{Soeding66,Drell60,Drell61},
which gives a non-resonant background.
The diagrams include the full $t$-channel scattering amplitudes for the
$KN \to KN$ and $K^*N \to KN$ scattering, but here we consider the tree
diagrams of one-meson exchange only.
For the intermediate $K\overline{K}$ pair, we consider the vector meson
exchanges.
The effective Lagrangians for defining these amplitudes are
\begin{eqnarray}
\mathcal{L}_{\gamma KK} &=& -ie A_\mu \left( K^- \partial^\mu K^+ -
\partial^\mu K^- K^+ \right),
\nonumber \\
\mathcal{L}_{VKK} &=& -i g_{VKK}^{} \left( \partial^\mu \overline{K}
V_\mu K - \overline{K} V_\mu \partial^\mu K \right),
\nonumber \\
\mathcal{L}_{VNN} &=& -g_{VNN}^{} \overline{N} \left\{ \gamma^\mu V_\mu -
\frac{\kappa_V}{2M_N} \sigma^{\mu\nu} \partial_\nu V_\mu \right\} N,
\nonumber \\
\mathcal{L}_{\gamma VKK} &=& -eg_{VKK}^{} \overline{K} \left[
(1+\tau_3)/2, V_\mu \right]_+ K A^\mu,
\label{eq:Lag1}
\end{eqnarray}
where $[A,B]_+ = AB+BA$, $V=(\rho,\omega,\phi)$, $A_\mu$ is the photon
field, and the kaon isodoublets are defined as
\begin{equation}
K = \left( \begin{array}{c} K^+ \\ K^0 \end{array} \right), \qquad
\overline{K} = \left( K^-, \overline{K}^0 \right).
\label{eq:k-isodoublet}
\end{equation}
We follow Ref.~\cite{SL96} to set%
\footnote{The Bonn potential gives $\kappa_\rho \approx
6.0$~\cite{MHE87}, which is larger than the value in Eq.~(\ref{gVN}).
We found that our results on the $\gamma N \to K \overline{K} N$ reaction
are not sensitive to this coupling constant.}
\begin{eqnarray}
&& g_{\rho NN}^{} = 3.1, \qquad \kappa_\rho = 2.0,  \nonumber \\
&& g_{\omega NN}^{} = 10.3, \qquad \kappa_\omega = 0.0.
\label{gVN}
\end{eqnarray}
The exchange of $\phi$ meson is neglected by taking the simplest OZI
rule prediction, $g_{\phi NN} = 0$.

We next need to define the vertices connecting the kaons and vector mesons
in Fig.~\ref{fig:Drell}.
This is done by using the following SU(3) symmetric Lagrangian,
\begin{eqnarray}
\mathcal{L}_{VPP} &=& -\frac{ig_{VPP}^{}}{\sqrt2} \mbox{ Tr} \{ V_\mu (P
\partial^\mu P - \partial^\mu PP) \},
\nonumber \\
\mathcal{L}_{VVP} &=& g_{VVP}^{} \varepsilon^{\mu\nu\alpha\beta}
\mbox{ Tr} (\partial_\mu V_\nu \partial_\alpha V_\beta P),
\label{eq:lag2}
\end{eqnarray}
where
\begin{equation}
P = \left( \begin{array}{cc}
 \frac{1}{\sqrt6} \eta + \frac{1}{\sqrt2} \bm{\pi} \cdot \bm{\tau} & K \\
\overline{K} & - \frac{2}{\sqrt6} \eta
\end{array} \right),
\qquad
V = \left( \begin{array}{cc}
 \frac{1}{\sqrt2} \omega + \frac{1}{\sqrt2} \bm{\rho} \cdot \bm{\tau} &
K^* \\
\overline{K}^* & - \phi
\end{array} \right).
\label{eq:octet}
\end{equation}
The $K^*$ isodoublets are defined by the same way as in
Eq.~(\ref{eq:k-isodoublet}).
The SU(3) symmetry relations lead to
\begin{equation}
g_{\rho KK}^{} = g_{\omega KK}^{} = g_{\rho\pi\pi}^{}/2 = 3.02.
\label{eq:VKK1}
\end{equation}

For the intermediate $K\overline{K}^*$ or $\overline{K}K^*$ pair
in Fig.~\ref{fig:Drell}, we consider the pseudoscalar meson exchanges.
The photon coupling is defined by
\begin{eqnarray}
\mathcal{L}_{\gamma K^*K} &=& g_{\gamma K^*K}^0
\epsilon^{\mu\nu\alpha\beta} \partial_\mu A_\nu \left( \partial_\alpha
K_\beta^{*0} \overline{K}^0 + \partial_\alpha K_\beta^{*0} K^{*0}
\right)
\nonumber \\
&& \mbox{} + g_{\gamma K^*K}^c
\epsilon^{\mu\nu\alpha\beta} \partial_\mu A_\nu \left( \partial_\alpha
K_\beta^{*-} K^+ + \partial_\alpha K_\beta^{*+} \overline{K}^- \right),
\label{eq:Lag2}
\end{eqnarray}
with
\begin{equation}
g_{\gamma K^*K}^0 = - 0.388 \mbox{ GeV}^{-1}, \qquad
g_{\gamma K^*K}^c = 0.254 \mbox{ GeV}^{-1},
\end{equation}
determined from the radiative decay widths of the neutral and charged $K^*$
vector mesons.
The couplings involving the pseudoscalar mesons $\pi$ and $\eta$ 
are defined by
\begin{eqnarray}
\mathcal{L}_{K^*K\varphi} &=& -ig_{K^*K\varphi}^{} \left( \overline{K}
\partial^\mu \varphi K_\mu^* - \overline{K}^*_\mu \partial^\mu \varphi K
\right), \nonumber \\
\mathcal{L}_{\varphi NN} &=& \frac{g_{\varphi NN}^{}}{2M_N} \overline{N}
\gamma^\mu \gamma_5 \partial_\mu \varphi N,
\end{eqnarray}
where $\varphi = \bm{\tau} \cdot \bm{\pi}, \eta$ and $N = (p,\ n)^T$.
We choose the usual $g_{\pi NN} = 13.4$ and use the SU(3) relation
to set $g_{\eta NN} = 3.54$.
By using the experimental value for $\Gamma(K^* \to K \pi)$ and
$\Gamma(K^* \to K \pi) = (g_{K^*K\pi}^2/8\pi M_{K^*}^2) p_\pi^3$,
we use
\begin{equation}
g_{K^* K \pi}^{} = 6.56.
\end{equation}
This value is close to the SU(3) value, $g_{K^* K \pi}^{} = g_{\rho\pi\pi}^{}
= 6.04$.
For $g_{K^* K \eta}^{}$, we use the SU(3) relation,
\begin{equation}
g_{K^*K\eta}^{} = \sqrt3 g_{K^* K \pi}^{} = 11.36
\end{equation}

In the diagrams of Fig.~\ref{fig:Drell}, the $K\overline{K}^*$ or
$\overline{K}K^*$ intermediate states can also interact with the nucleon
via vector meson exchanges.
This can be calculated from the Lagrangian derived from $\mathcal{L}_{VVP}$
of Eq.~(\ref{eq:lag2}),
\begin{equation}
\mathcal{L}_{K^*KV} = g_{K^*KV}^{} \varepsilon^{\mu\nu\alpha\beta}
\overline{K} \partial_\mu V_\nu \partial_\alpha K_\beta^* + \mbox{
H.c.},
\end{equation}
where
\begin{equation}
g_{K^*K\rho}^{} = g_{K^*K\omega}^{} = g_{\omega\rho\pi}/2, \qquad
g_{K^*K\phi} = g_{\omega\rho\pi}/\sqrt2.
\end{equation}
The above coupling constants can then be fixed 
by using the hidden gauge approach \cite{BKY88} to set
\begin{equation}
g_{\omega\rho\pi} = \frac{N_c g_\rho^2}{8\pi^2 f_\pi} = 14.9
\mbox{ GeV}^{-1},
\end{equation}
where $N_c = 3$, $g_\rho = g_{\rho\pi\pi}^{}$, and $f_\pi = 93$ MeV.

\subsection{Vector meson background}

\begin{figure}[t]
\centering
\epsfig{file=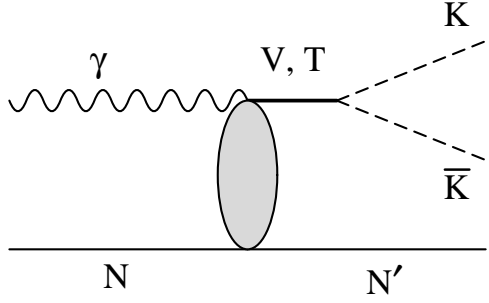, width=0.4\hsize}
\caption{Vector and tensor meson photoproduction contributions to
$\gamma N \to K \overline{K} N$. Here $V$ stands for a vector meson
($V=\rho,\omega,\phi$) and $T$ for a tensor meson ($T=a_2^{},f_2$).}
\label{fig:VTM}
\end{figure}

As shown in Fig.~\ref{fig:VTM}, the vector meson background amplitude
is determined by a vector meson photoproduction amplitude and the
$V \to K\overline{K}$ vertex function defined by $\mathcal{L}_{VKK}$
of Eq.~(\ref{eq:Lag1}).
The resulting $\gamma N \to K \overline{K} N$ amplitude can be written as
\begin{equation}
\mathcal{M} = \overline{u}_{N'}^{}(p') 
\mathcal{M}^{\mu} \varepsilon_\mu u_N^{}(p),
\label{eq:v-m}
\end{equation}
where $u_{N}^{}(p)$ is the Dirac spinor of a nucleon with four-momentum $p$,
$\varepsilon_\mu$ is the photon polarization vector, $p_\mu$ and $p_\mu'$
are the four-momenta of the initial and final nucleon, respectively.
The dynamics of Eq.~(\ref{eq:v-m}) is contained in the following invariant
amplitude,
\begin{equation}
\mathcal{M}^\mu = \mathcal{M}^{\nu\mu}(\gamma N \to VN)
\frac{g_{VKK}^{}}{(q_1+q_2)^2 - M_V^2 } (q_1-q_2)_\nu
F_V^{(i)},
\label{eq:vm-part}
\end{equation}
where $q_1$ and $q_2$ are the momenta of the outgoing $K$ and $\overline{K}$,
$F_V^{(i)}$ depends on the channel quantum number $i$ as well as a form
factor (\ref{eq:ff}) which takes into account the off-shell--ness of the
intermediate vector meson.
The needed coupling constants $g_{VKK}^{}$ with $V=\rho,\omega$ have been
given in Eq.~(\ref{eq:VKK1}).
For $\phi$ decay, we here use $g_{\phi KK} =-4.49 $ deduced from the
experimental data of $\Gamma(\phi \rightarrow K^+K^-)$~\cite{PDG04},
while the phase is taken from the SU(3) symmetry.
The photoproduction amplitude $\mathcal{M}^{\nu\mu}(\gamma N \to VN)$ is
generated from a phenomenological model within which the blob in
Fig.~\ref{fig:VTM} includes Pomeron exchange, $\pi$, $f_2$, and other
meson exchanges, and the direct and crossed nucleon terms.
The details of this model can be found in Refs.~\cite{OTL00,OTL01,OL02,OL04}
and will not be repeated here.
But it should be mentioned that those models describe well the
experimental data for vector meson photoproduction.
The finite decay width of vector mesons is included by replacing $M_V$
by $M_V - i \Gamma_V/2$ in Eq.~(\ref{eq:vm-part}).

\subsection{Tensor meson background}

The tensor meson photoproduction contribution to $\gamma N \to
K\overline{K}N$ is particularly interesting since it was suggested in
Ref.~\cite{DKST03} that this mechanism can generate a peak near 1540 MeV
in the $KN$ invariant mass distribution of the $\gamma n \to K^+K^- n$
reaction and hence the discovery of $\Theta(1540)$ pentaquark baryon is
questionable.
In this subsection, we explore this mechanism in more detail.

As illustrated in Fig.~\ref{fig:VTM}, the tensor meson photoproduction
contribution is very similar to the vector meson photoproduction contribution.
Its contribution to the $\gamma N \to K\overline{K}N$ amplitude is of
the same structure of Eqs.~(\ref{eq:v-m})-(\ref{eq:vm-part}) and can be
written as
\begin{equation}
\mathcal{M} = \overline{u}_{N'}^{}(p') 
\mathcal{M}^{\mu} \varepsilon_\mu u_N^{}(p).
\label{eq:t-m}
\end{equation}
The main dynamics of Eq.~(\ref{eq:t-m}) is contained in
\begin{eqnarray}
\mathcal{M}^\mu &=& 
 \mathcal{M}^{\gamma\delta,\mu}(\gamma N \to TN)
\frac{2G_{TKK}^{}}{M_T} \frac{P_{\gamma\delta;\rho\sigma}}{(q_1+q_2)^2
- M_T^2 } q_1^\rho q_2^\sigma F_T^{(i)},
\label{eq:tm-part}
\end{eqnarray}
where $q_1$ and $q_2$ are the momenta of the outgoing $K$ and $\overline{K}$.
$F_T^{(i)}$ includes the constant depending on the channel $i$ and the
form factor.
Here the $T\to K\overline{K}$ decay vertex is defined by the following
tensor structure associated with a spin $J=2$ particle, whose propagator
contains
\begin{equation}
P^{\mu\nu;\rho\sigma} = \frac12 \left(\overline{g}^{\mu\rho}
\overline{g}^{\nu\sigma} + \overline{g}^{\mu\sigma}
\overline{g}^{\nu\rho} \right) - \frac13 \overline{g}^{\mu\nu}
\overline{g}^{\rho\sigma},
\end{equation}
with
\begin{equation}
\overline{g}^{\mu\nu} = -g^{\mu\nu} + \frac{1}{M_T^2} p^\mu p^\nu,
\end{equation}
where $p^\mu$ is the momentum of the tensor meson.
The coupling constant $G_{TKK}$ in Eq.~(\ref{eq:tm-part}) is defined by
the Lagrangian
\begin{equation}
\mathcal{L}_{TKK}^{} = - \frac{2G_{TKK}}{M_T^{}} \partial_\mu \overline{K}
\, T^{\mu\nu} \, \partial_\nu K,
\label{eq:TKK}
\end{equation}
where the kaon isodoublets are defined in Eq.~(\ref{eq:k-isodoublet})
and $T^{\mu\nu} = f^{\mu\nu}$ or $\bm{\tau} \cdot \bm{a}^{\mu\nu}$,
with $f^{\mu\nu}$ and $a^{\mu\nu}$ denoting the isoscalar $f_2(1275)$ and 
isovector $a_2(1320)$ tensor meson field, respectively.
Some details about $\mathcal{L}_{TKK}^{}$ can be found in Appendix A.
Using the experimental data~\cite{PDG04},
$\Gamma(f_2 \to K\overline{K})_{\rm expt.} \approx 8.6$ MeV and
$\Gamma(a_2 \to K\overline{K})_{\rm expt.} \approx 5.24$ MeV,
Eq.~(\ref{eq:TKK}) leads to
\begin{equation}
G_{fKK} = 7.15, \qquad G_{aKK} = 4.89.
\end{equation}
In this work, we neglect the contribution from $f_2'(1525)$ meson since
it can be produced favorably only at energies much higher than the region
considered in this work.
The finite decay width of tensor mesons is included by replacing $M_T$
by $M_T - i \Gamma_T/2$ in its propagator.

We now turn to discussing the calculations of the tensor meson photoproduction 
amplitude $\mathcal{M}^{\alpha\beta,\mu}$ of Eq.~(\ref{eq:tm-part}).
We note here that the tensor meson photoproduction mechanisms depend very
much on the charge of the produced tensor mesons.
In particular, the one-pion exchange is known~\cite{EHHP69,EHRS72,CHBB93,AS00}
to be the dominant mechanism for charged tensor meson production, while it is
not allowed in neutral tensor meson production because of $C$ parity.
Thus the claim made by Dzierba {\it et al.\/}~\cite{DKST03} concerning the
peak at $W\sim 1540$ MeV generated by neutral tensor meson production must be
re-examined.
This will be our focus by exploring the contributions from the vector meson
exchange mechanisms.

\subsubsection{charged tensor meson photoproduction}

\begin{figure}[t]
\centering
\epsfig{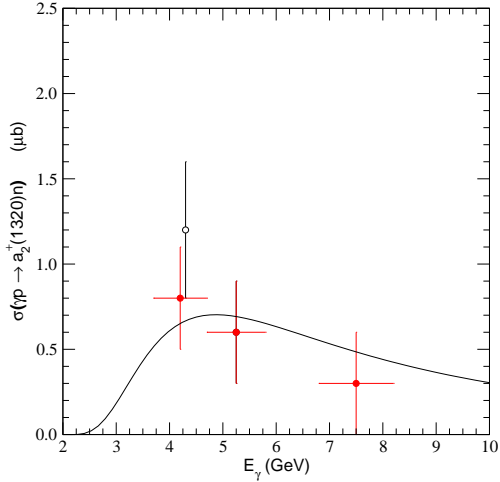}
\caption{
Total cross section for charged tensor meson photoproduction,
$\gamma p \to a_2^+(1320)n$. Experimental data are from Refs.
\cite{EHHP69} ($\circ$) and \cite{EHRS72} ($\bullet$).}
\label{fig:a2mp}
\end{figure}

We first calculate the charged tensor meson photoproduction,
$\gamma p \to a_2^+(1320) n$ to explore the one-pion exchange model.
This can be done by using the following interaction
Lagrangian~\cite{HHKS66} which defines the $a_2\gamma\pi$ coupling,
\begin{equation}
\mathcal{L}_{a_2\gamma\pi} = \frac{g_{a_2\gamma\pi}}{M_a^2}
\varepsilon^{\mu\nu\alpha\beta} \partial_\mu A_\nu a^\pm_{\alpha\lambda}
(\partial^\lambda \partial_\beta \pi^\mp).
\label{eq:Lag_a2gp}
\end{equation}
The decay width $\Gamma(a_2^\pm \to \pi^\pm \gamma)$ then reads
\begin{equation}
\Gamma(a_2^\pm \to \pi^\pm \gamma) = \frac{g_{a_2\gamma\pi}^2}{40\pi}
\frac{p_\gamma^5}{M_a^4},
\end{equation}
where
$p_\gamma^{} = ( M_a^2 - M_\pi^2 )/2M_a$.
Using $\Gamma(a_2 \to \gamma\pi)_{\rm expt.} \approx 0.29$ MeV, we get
\begin{equation}
g_{a_2\gamma\pi} \approx 0.96.
\end{equation}
Then the $\gamma p \to a_2^+(1320) n$ amplitude due to one-pion-exchange
is obtained as
\begin{eqnarray}
\mathcal{M}^{\mu\nu,\alpha} &=& \frac{\sqrt2 g_{\pi NN}
g_{a_2^{}\gamma\pi}}{M_a^2} \frac{1}{(p-p')^2 - M_\pi^2}
\varepsilon^{\rho\alpha\mu\sigma} k_\rho (q-k)^\nu (q-k)_\sigma
\gamma_5 F\bm{(}M_\pi,(k-q)^2)\bm{)},
\end{eqnarray}
where $k$ is the photon momentum and $q$ is the tensor meson momentum.
The form factor is introduced in the form of
\begin{equation}
F(M,r) = \left(\frac{\Lambda^4}{\Lambda^4 + (r-M^2)^2} \right)^2.
\label{eq:ff}
\end{equation}

We adjust the cutoff $\Lambda$ of the form factor to fit the total
cross section of $\gamma p \to a_2^+(1320)n$.
With $\Lambda = 0.45$ GeV, the result is shown in Fig.~\ref{fig:a2mp}.
Therefore, we confirm that the photoproduction of charged $a_2$ meson
can be reasonably described by the
one-pion-exchange~\cite{EHHP69,EHRS72,CHBB93,AS00}.

\subsubsection{neutral tensor meson photoproduction}

In Ref.~\cite{DKST03}, the authors used one-pion-exchange for neutral
tensor meson photoproduction by extending the model for charged tensor
meson photoproduction.
However, the one-pion-exchange is {\em not allowed\/} for neutral tensor meson
photoproduction because of the $C$-parity.
Instead, we expect that the vector-meson exchange is the dominant process at
low energies since the lightest mesons with $C=-1$ are the neutral vector
mesons, $\rho^0$ and $\omega$.
At high energies, the Odderon exchange, a partner of the Pomeron with
odd $C$ parity, is suggested as the major production mechanism and in
fact neutral tensor meson photoproduction process has been suggested to
study the Odderon exchange \cite{BDDN00}.
Since the role of the Odderon exchange is not clearly known, especially
at low energies, we will mainly consider
 the vector-meson exchange for the production
mechanism of neutral tensor meson photoproduction in the energy region
of our interest, i.e., $E_\gamma \le 3$ GeV.

\begin{table}[t]
\centering
\begin{tabular}{c|cc} \hline\hline
Decay & Decay width (keV) & $g_{TV\gamma}^{}$ \\ \hline
$f_2 \to \rho\gamma$ & $254$ & $0.14$ \\
$f_2 \to \omega\gamma$ & $27$ & $0.048$ \\
$f_2 \to \phi\gamma$ & $1.3$ & $-0.022$ \\
$f_2' \to \rho\gamma$ & $4.8$ & $0.0145$ \\
$f_2' \to \omega\gamma$ & $\sim 0$ & $\sim 0$ \\
$f_2' \to \phi\gamma$ & $104$ & $0.10$  \\ \hline
$a_2 \to \rho\gamma$ &  $28$ & $0.044$ \\
$a_2 \to \omega\gamma$ & $247$ & $0.134$ \\
$a_2 \to \phi\gamma$ & $0.8$ & $-0.015$
\\ \hline\hline
\end{tabular}
\caption{Decay widths and couplings for $T \to V\gamma$ decay in a
covariant quark model~\cite{IYO89}.}
\label{tab:TVG}
\end{table}

To calculate the vector meson exchange amplitude, we need to first
define the $TV\gamma$ coupling, where $T=f_2, a_2$ and
$V=\rho, \omega, \phi$.
This can be determined by considering the tensor meson decay amplitude which
can be written in the most general form as~\cite{Renn71}
\begin{equation}
\langle \gamma(k) V(k') | T \rangle = \frac{1}{M_T} \epsilon^\kappa
\epsilon'^\lambda \epsilon^{\mu\nu} A_{\kappa\lambda\mu\nu}(k,k'),
\end{equation}
where
\begin{eqnarray}
A^{}_{\kappa\lambda\mu\nu} (k,k') &=&
\frac{f^{}_{TV\gamma}}{M_T^3} \left[ g_{\kappa \lambda} (k\cdot k')
- k'_\kappa k_\lambda \right] (k-k')_\mu (k-k')_\nu ]
\nonumber \\ && \mbox{} 
+ g_{TV\gamma}^{} [ g_{\kappa\lambda} (k-k')_\mu (k-k')_\nu
+ g_{\lambda\mu} k'_\kappa (k-k')_\nu
+ g_{\lambda\nu} k'_\kappa (k-k')_\mu
\nonumber \\ && \mbox{} \qquad
- g_{\kappa\mu} k_\lambda (k-k')_\nu
- g_{\kappa\nu} k_\lambda (k-k')_\mu
-2 k \cdot k' ( g_{\kappa\mu} g_{\lambda\nu} + g_{\kappa\nu}
g_{\lambda\mu}) ].
\label{eq:TVG}
\end{eqnarray}
The above form is known to give better descriptions of the known tensor meson
radiative decays~\cite{Singer83}.
For simplicity, we will take the assumption of tensor meson dominance which
leads to $f_{TV\gamma}^{} \approx 0$ \cite{Renn71}.
Without the $f_{TV\gamma}$ term, Eq.~(\ref{eq:TVG}) is equivalent
to that of Ref.~\cite{TEHN04} except for the factor of 2 difference in the
definition of the coupling constant $g_{TV\gamma}^{}$.
In the present calculation, we take $g_{TV\gamma}^{}$ from the covariant
quark model predictions~\cite{IYO89}, which gives reasonable description of
the known radiative decay widths of vector and tensor mesons.
These values are listed in Table~\ref{tab:TVG} along with the predicted
decay widths.
The details on the calculation of the radiative decays of tensor mesons are
given in Appendix A.

\begin{figure}[t]
\centering
\epsfig{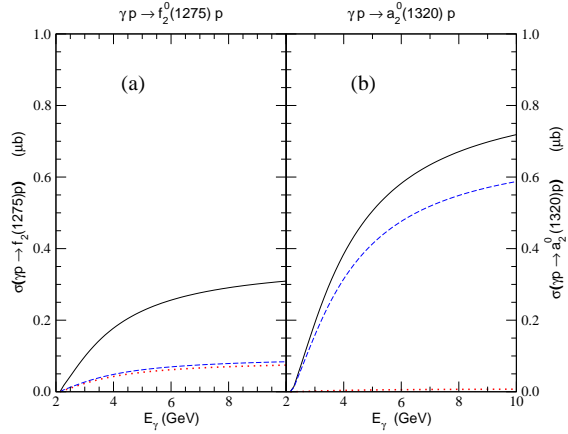}
\caption{
Total cross section for neutral tensor meson photoproduction,
(a) $\gamma p \to f_2^0(1275)p$ and (b) $\gamma p \to a_2^0(1320)p$.
The dotted lines are from $\rho$ meson exchange and the dashed lines
from $\omega$ meson exchange. The solid lines are their sums.}
\label{fig:f2a20}
\end{figure}

With the $TV\gamma$ coupling fixed and the $VNN$ coupling defined
in the previous subsection, we can write the $\gamma p \rightarrow T p$
amplitude as
\begin{eqnarray}
\mathcal{M}^{\mu\nu,\alpha} &=& - \frac{1}{M_T}
A^{\alpha\beta\mu\nu}(k,q-k) \frac{C_{VNN} g_{VNN}}{(p-p')^2 - M_V^2}
\left\{ g_{\beta\delta} - \frac{1}{M_V^2} (p-p')_\beta (p-p')_\delta
\right\}
\nonumber \\ && \mbox{} \times
\left\{ \gamma^\delta - i \frac{\kappa_V}{2M_N} \sigma^{\delta\lambda}
(p-p')_\lambda \right\},
\end{eqnarray}
with $C_{VNN} = -1$ for $\rho nn$, and $C_{VNN} = 1$ for $\rho pp$,
$\omega pp$, and $\omega nn$.
The form of $A^{\alpha\beta\mu\nu}(k,q-k)$ is defined by Eq.~(\ref{eq:TVG}).
The form factor (\ref{eq:ff}) is also used here to regularize each vertex.
However, the experimental data for neutral tensor meson photoproduction is
very limited and uncertain and cannot be used to fix the cutoff parameter
$\Lambda$.%
\footnote{In this work, we do not use the data for the backward
scattering cross sections for $f_2$ meson photoproduction of
Ref.~\cite{AHHM76}, which was obtained by hand-drawn curves and after
corrections for unobserved decay modes.}
To estimate their contribution, we take a relatively large cutoff
$\Lambda = 0.9$ GeV to calculate the total cross sections for $f^0_2(1275)$
and $a^0_2(1320)$ photoproduction. 
The results are shown in Fig.~\ref{fig:f2a20}.
We see that they are smaller than the charged tensor meson production cross
sections shown in Fig.~\ref{fig:a2mp} when $E_\gamma \le 5$ GeV.
If a smaller cutoff such as $\Lambda=0.45$ GeV employed in
Fig.~\ref{fig:a2mp} is used, the predicted cross sections will be even
smaller.
We will use $\Lambda=0.9$ GeV in our calculation as an estimate of an upper
bound of the neutral tensor meson photoproduction contribution.
This will allow us to examine whether a peak near 1540 MeV in $KN$ invariant
mass can be generated by neutral tensor meson photoproduction process in the
$\gamma n \rightarrow K^+K^- n$ reaction.

{}From our results, we also found that in the case of $a_2^0$
photoproduction, the $\omega$ meson exchange is dominant, while both the
$\omega$ and $\rho$ exchanges are comparable in $f_2^0$ photoproduction.
This is because $g_{a_2\rho\gamma}/g_{a_2\omega\gamma} \sim 1/3$
and $g_{f_2\rho\gamma}/g_{f_2\omega\gamma} \sim 3$ as seen in
Table~\ref{tab:TVG}, while $g_{\rho^0 NN}/g_{\omega NN} \sim 1/3$.
Thus the $\rho$ meson exchange amplitude in $a_2^0$ photoproduction is
suppressed by an order of magnitude than the $\omega$ meson exchange,
while the $\rho$ and $\omega$ meson exchanges have similar magnitude
in $f_2$ photoproduction.%
\footnote{Therefore, our results are different from those of
Ref.~\cite{TEHN04}, where only the $\omega$ exchange for $a_2^0$
photoproduction and only the $\rho$ exchange for $f_2$ photoproduction were
considered. We also calculated neutral tensor meson photoproduction
in Regge model by Reggeizing the vector meson exchange amplitudes.
The obtained results show that the total cross section decreases with energy
as expected, but we found that the maximum values obtained in Regge model
without form factor are close to the ones given in Fig.~\ref{fig:f2a20}
with the similar photon energy.}

\subsection{Hyperon background}

\begin{figure}[t]
\centering
\epsfig{file=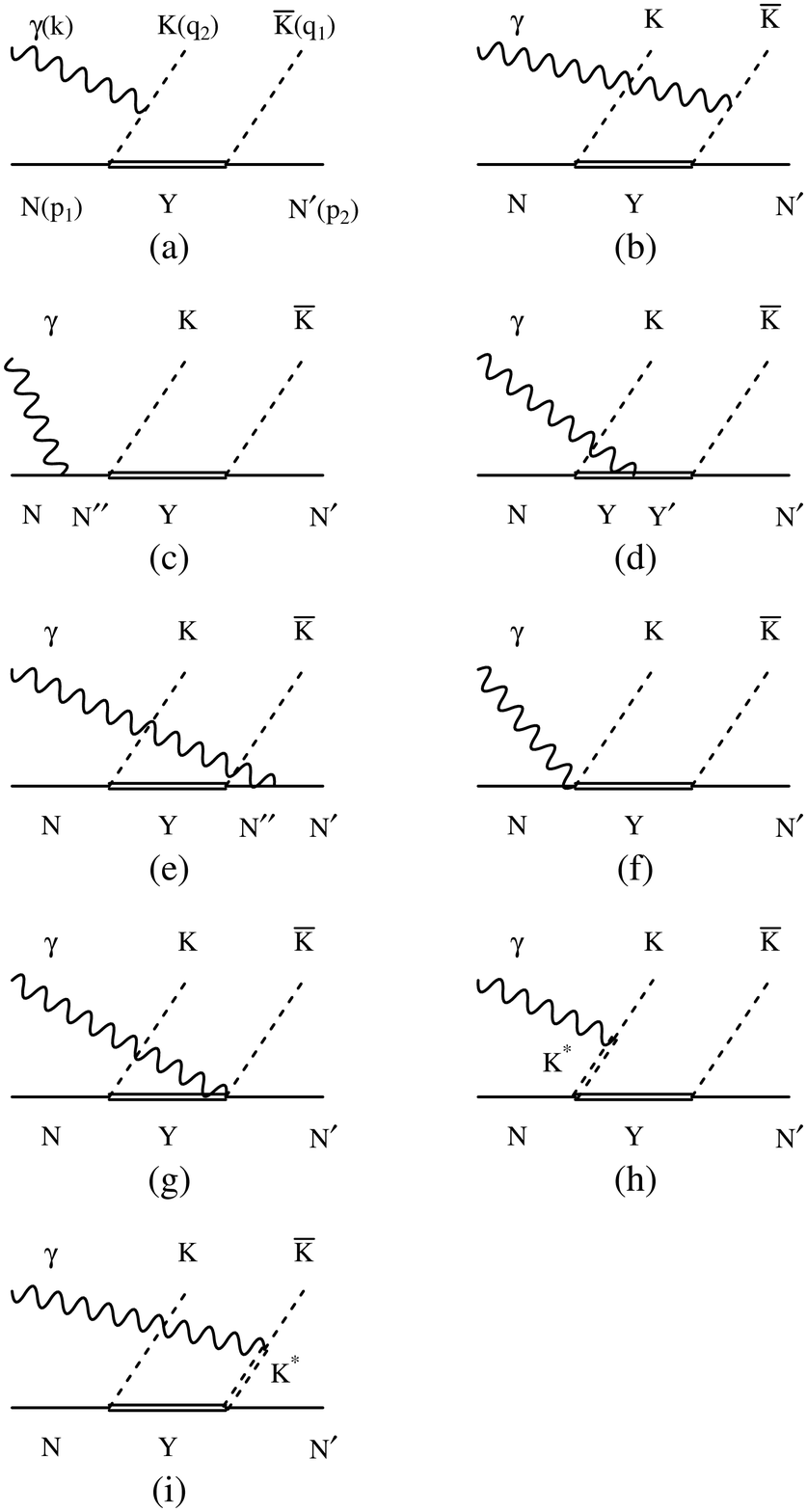, width=0.45\hsize}
\caption{$S=-1$ hyperon backgrounds to $\gamma N \to K\overline{K} N$.}
\label{fig:hyp}
\end{figure}

The hyperon backgrounds for $\gamma N \to K\overline{K}N$ were considered in
Refs.~\cite{NT03,Roberts04}.
The main difficulty in estimating the hyperon background arises from the
uncertainties of the coupling parameters of the hyperon resonances
($Y=\Lambda, \Sigma$).
We, therefore, consider only the well-known hyperon resonances,
i.e., $\Lambda(1116)$, $\Lambda(1405)$, $\Lambda(1520)$, $\Sigma(1193)$,
and $\Sigma(1385)$.
We take the pseudoscalar coupling of kaons with the spin $J=1/2$ hyperons.
For coupling with the spin $J=3/2$ hyperons, we follow the Rarita-Schwinger
formulation of Refs.~\cite{ELP92,LLD91,DMW91,BDM89,BMZ95}.
The resulting hyperon background tree-diagrams are shown in Fig.~\ref{fig:hyp}.
We note that Fig.~\ref{fig:hyp}(d) includes the photo-transitions among the
hyperons, which were not included in the previous works.
Furthermore, Figs.~\ref{fig:hyp}(c) and (e) contain strangeness $S=0$ state
$N''$. 
We will only consider the possibilities that $N''$ is either the nucleon
($N$) or the Delta [$\Delta (1232)$], since the experimental information
about the transitions from other higher mass nucleon resonances to hyperons
is not well-known.
Since we are using the pseudoscalar coupling for spin-1/2 baryons, there
is no $\gamma K YN$ contact interaction and hence we have five diagrams,
Figs.~\ref{fig:hyp}(a)-(e) for the intermediate $\Lambda(1116)$,
$\Lambda(1405)$, and $\Sigma(1193)$ when $K^*$ intermediate state is neglected.
But for spin-3/2 $\Lambda(1520)$ and $\Sigma(1385)$, the $\gamma K YN$ contact
interaction is induced and we have seven diagrams, namely,
Figs.~\ref{fig:hyp}(a)-(g) in the absence of $K^*$ intermediate state.
Figures~\ref{fig:hyp}(h)-(i) account for the effects due to $K^*$
intermediate state.

The Lagrangians defining the photon coupling with $K$ and $K^*$
in Figs.~\ref{fig:hyp}(a), (b), (h), and (i) have been specified in
Eqs.~(\ref{eq:Lag1}) and (\ref{eq:Lag2}).
The other Lagrangians used in the calculation of the diagrams in
Fig.~\ref{fig:hyp} are given in Appendix B in detail.
There, we also discuss how the other couplings are defined by using SU(3)
symmetry, experimental information, and some hadron model predictions.

\subsection{Pentaquark contribution}

\begin{figure}[t]
\centering
\epsfig{file=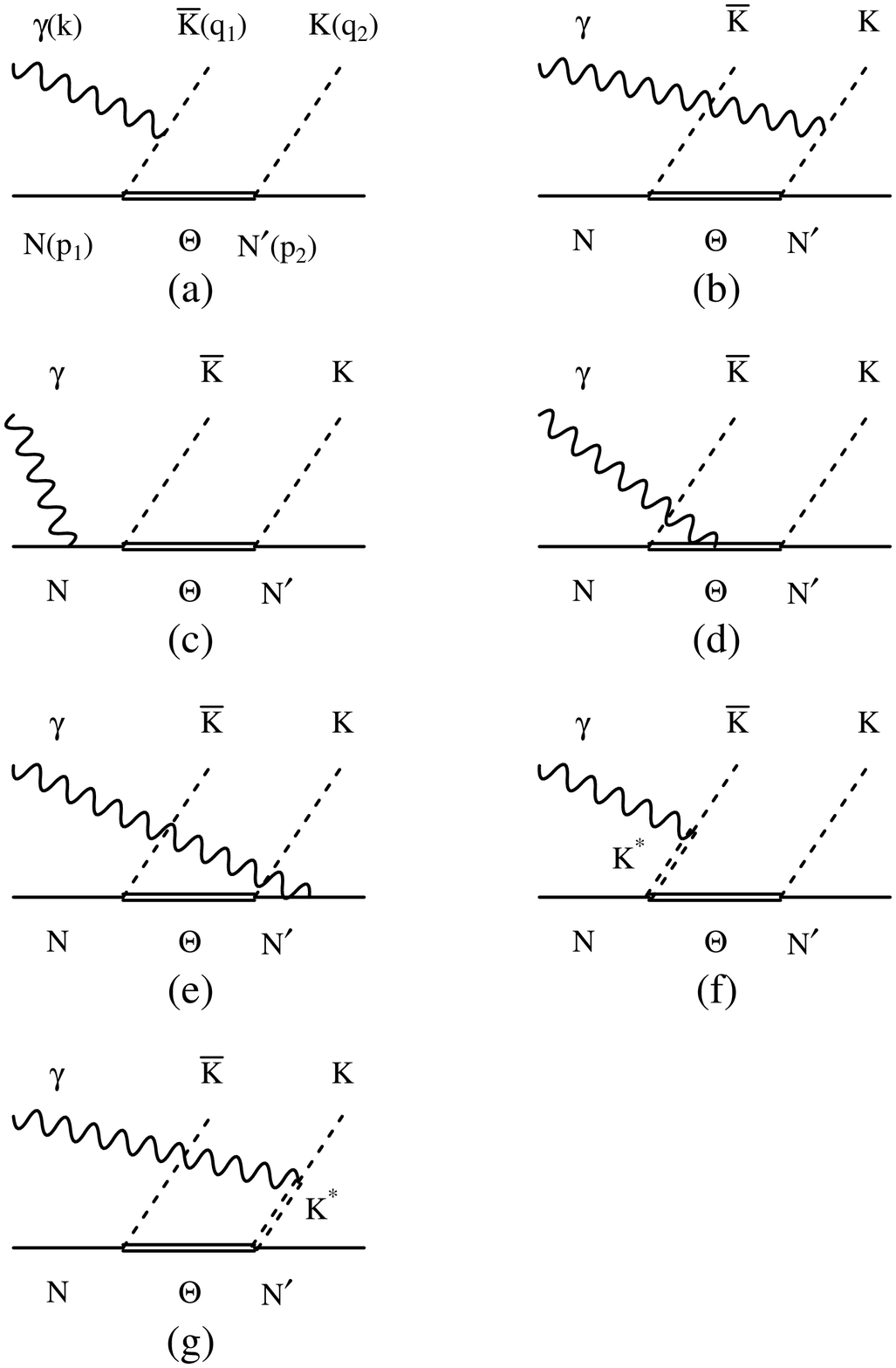, width=0.45\hsize}
\caption{Exotic $S=+1$ pentaquark $\Theta$ contribution to
$\gamma N \to K\overline{K} N$.}
\label{fig:theta}
\end{figure}

The pentaquark $\Theta(1540)$ contribution to $\gamma N \to
K\overline{K}N$ is depicted in Fig.~\ref{fig:theta}.
We first assume that it belongs to pentaquark baryon antidecuplet,
which means that the $\Theta(1540)$ is an isosinglet state.
In this case, $\Theta(1540)$ can contribute to, or can be observed in the
reactions of $\gamma p \to K^0\overline{K}^0 p$ and $\gamma n \to K^+
K^- n$.
There is still no experimental clue on the existence of isovector
$\Theta$ or isotensor $\Theta$, which are members of pentaquark
$\bm{27}$-plet and $\bm{35}$-plet, respectively.
So we do not consider such cases and focus on the isosinglet $\Theta^+$
since the nonexistence of a peak at 1540 MeV in $K^+ p$ channel strongly
suggests the isosinglet nature of $\Theta(1540)$, if exists.
Next we assume that $\Theta(1540)$ has spin-1/2 following
most hadron model predictions.
There may be its higher spin resonances, but it will not be investigated
in this study.
However, there are strong debates on the parity of $\Theta(1540)$.
Therefore, in order to study the sensitivity of the physical quantities
in $\gamma N \to K\overline{K}N$ on the parity of $\Theta(1540)$, we
allow the both parities.
Thus the effective Lagrangians are
\begin{eqnarray}
\mathcal{L}_{KN\Theta} &=& -ig_{KN\Theta}^{}
\overline{\Theta} \Gamma^\pm \overline{K}^c N + \mbox{ H.c.},
\nonumber \\
\mathcal{L}_{K^*N\Theta} &=& - g_{K^*N\Theta}^{} \overline{\Theta} \left(
\Gamma_\mu^\pm \overline{K}^{*c\mu} - \frac{\kappa^T_{K^*N\Theta}}
{M_N + M_\Theta} \Gamma^\mp \sigma^{\mu\nu} \partial_\nu
\overline{K}_\mu^{*c} \right) N
+ \mbox{ H.c.},
\nonumber \\
\mathcal{L}_{\gamma\Theta\Theta} &=& -e \overline{\Theta} \left[ A_\mu
\gamma^\mu - \frac{\kappa_\Theta^{}}{2M_N^{}} \sigma_{\mu\nu}
\partial^\nu A^\mu \right] \Theta,
\end{eqnarray}
where
\begin{equation}
\Gamma^\pm = \left( \begin{array}{c} \gamma_5 \\ 1 \end{array}
\right), \qquad \Gamma_\mu^\pm = \left( \begin{array}{c} \gamma_\mu \\
\gamma_\mu \gamma_5 \end{array} \right),
\end{equation}
and
\begin{equation}
K^c = \left( \begin{array}{c} - \overline{K}^0 \\ K^- \end{array} \right),
\qquad
K^{*c} = \left( \begin{array}{c} - \overline{K}^{*0} \\ K^{*-} \end{array}
\right).
\end{equation}
Here, the upper components of $\Gamma^\pm$ and $\Gamma_\mu^\pm$ are
for the even-parity $\Theta^+$ and the lower components for the odd-parity
$\Theta^+$.

For the couplings of $\Theta(1540)$, we assume that
$\Gamma(\Theta) = 1$ MeV \cite{ASW03a,ASW03b}, which is close to the value of
Particle Data Group \cite{PDG04}, $0.9 \pm 0.3$ MeV.
We then have
\begin{equation}
g_{KN\Theta}^{} = 0.984 \ (0.137)
\end{equation}
for positive (negative) parity $\Theta$.
There is no information about the $K^*N\Theta$ couplings.
As in Ref.~\cite{OKL03d}, we use $g_{K^*N\Theta}/g_{KN\Theta} = \sqrt3$
for even parity $\Theta$ and $g_{K^*N\Theta}/g_{KN\Theta} = 1/\sqrt3$
for odd parity $\Theta$ following the quark model
predictions~\cite{CD04,CCKN04b,CCKN03c} while neglecting the tensor coupling
terms.
The dependence of our results on this ratio will be discussed later.
As for the magnetic moment of $\Theta^+(1540)$, we use the prediction from
Ref.~\cite{BGS04}, $\mu(\Theta) \approx 0.1$ ($0.4$), which gives
$\kappa_\Theta \approx -0.9$ ($-0.6$) for positive (negative) parity
$\Theta$.
However, we found that our results are not sensitive to the value of
$\kappa_\Theta$.

\section{Results}

Like all of the effective Lagrangian approaches, the considered
tree-diagrams shown in the previous Section need to be regularized
by introducing form factors.
In this work, we use the form (\ref{eq:ff}) which was already used in our
investigation of tensor meson photoproduction in Section {II.1}.
Namely, a form factor of the form of Eq.~(\ref{eq:ff}) is introduced at
each vertex for all tree-diagrams in Section II, where $M$ is the mass of
the exchanged (off-shell) particle and $r$ is its four-momentum squared. 
Therefore, when it is on its mass-shell ($r=M^2$), the form factor
becomes $1$.

It is well-known that introducing form factors breaks the charge
conservation condition.
In order to satisfy this constraint, we extend the methods of
Refs.~\cite{Ohta89,Habe97,HBMF98a,DW01a}.
Namely, we introduce contact diagrams to restore the current conservation.
The details on this procedure are given and discussed in Appendix C.
Some of the cutoff parameters $\Lambda$ are already fixed by the available
experimental data for meson photoproduction as discussed in the previous
Section.
For simplicity, we set the remaining cutoff parameters the same for
all tree-diagrams and adjust it to fit the available experimental data of
$\gamma N \to K\overline{K}N$ reactions.
Obviously, it is not easy to interpret the resulting form factor theoretically.
Rather it should be just considered as a part of our phenomenological approach.

\subsection{Total cross sections}

\begin{figure}[t]
\centering
\epsfig{file=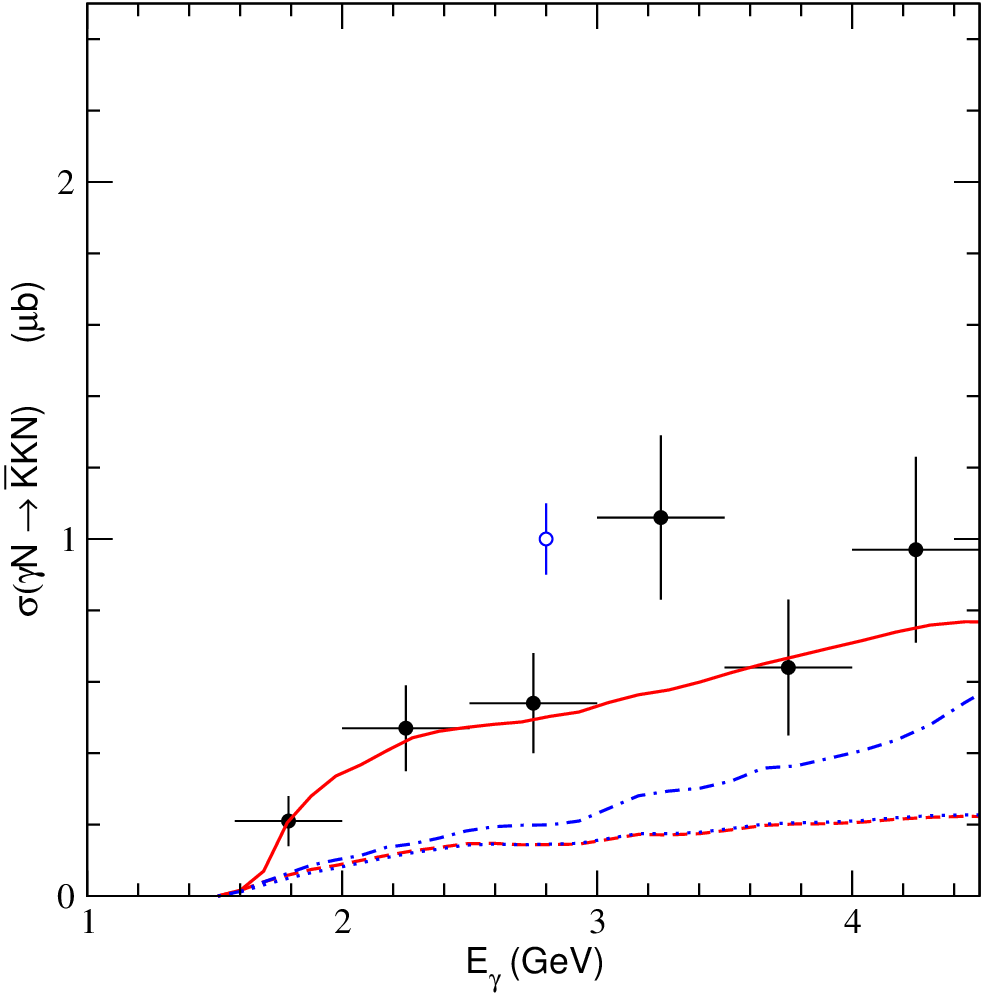, width=0.4\hsize}
\caption{
Total cross sections for $\gamma N \to K\overline{K} N$
reactions. The solid line is for $\gamma p \to K^+ K^- p$, dashed line
for $\gamma p \to K^0 \overline{K}^0 p$, dot-dashed line for $\gamma n
\to K^+ K^- n$, and dotted line for $\gamma n \to K^0 \overline{K}^0 n$.
The experimental data are for $\gamma p \to K^+ K^- p$ reaction and from
Ref.~\cite{ABBH69} $(\bullet)$ and Ref.~\cite{BCGG72} $(\circ)$.
The dashed and dotted lines are close together and hard to be
distinguished.}
\label{fig:total}
\end{figure}

Since we assume that the $\Theta(1540)$ is a particle with isospin $I=0$,
strangeness $S=+1$ and charge $Q=+|e|$, the $\Theta$ production
mechanisms (Fig.~\ref{fig:theta}) can not take place in the
$\gamma p \to K^+K^-p$  reaction.
We thus determine the cutoff parameter $\Lambda$ of the form factors by
fitting the available total cross section data for $\gamma p \to K^+ K^- p$.
In this way the background (non-$\Theta$ mechanism) amplitude can be fixed,
such that the identification of the $\Theta$ from the available data of 
$\gamma n \to K^+ K^- n$ reaction can be assessed.
With $\Lambda = 0.9$ GeV, our fit (solid curve) is shown in
Fig.~\ref{fig:total}.
Clearly, the data can be reproduced reasonably well.
With the same cutoff parameter, we then predict the total cross sections
for the other three processes listed in Eq.~(\ref{eq:channel}).
They are also shown in Fig.~\ref{fig:total}:
$\gamma p \to K^0 \overline{K}^0 p$ (dashed line), 
$\gamma n \to K^+ K^- n$ (dot-dashed line),
and $\gamma n \to K^0 \overline{K}^0 n$ (dotted line).
We can see that the cross sections for neutral $K^0\overline{K}^0$ pair
photoproduction from $p$ (dashed curve) or $n$ (dotted curve) 
are almost indistinguishable, and the cross sections for charged $K^+K^-$
pair photoproduction (solid and dash-dot curves) are larger than those for
neutral $K^0\overline{K}^0$ pair production.
The contributions from the $\Theta(1540)$ are not visible in the
calculated total cross sections.
The $\Theta$ can be identified only in the invariant mass distributions.

\subsection{Invariant mass distributions}

With the cutoff $\Lambda = 0.9$ GeV determined from the total cross
section for the $\gamma p \to K^+ K^-p$ reaction, we can now compare our
predictions of the invariant mass distributions for the
$\gamma n \to K^+ K^- n$ reaction with the JLab data~\cite{CLAS03-b}.
Unfortunately, the JLab data are not scaled properly for comparing
with the absolute magnitudes of the predicted cross sections because the
JLab data just show the invariant mass distributions in an arbitrary unit.
We therefore simply scale the JLab data to see whether we can roughly
reproduce the shape of the data within our model.%
\footnote{It should also be mentioned that the JLab data were not obtained
with a single photon energy, instead the data were taken by incident
electrons of $2.474$ and $3.115$ GeV, and the nuclear effects such as
final state interactions are not properly taken into account.}
This must be a very crude assumption but will be enough to study the
shape, especially the $\Theta$ peak, of the data.

\begin{figure}[t]
\centering
\epsfig{file=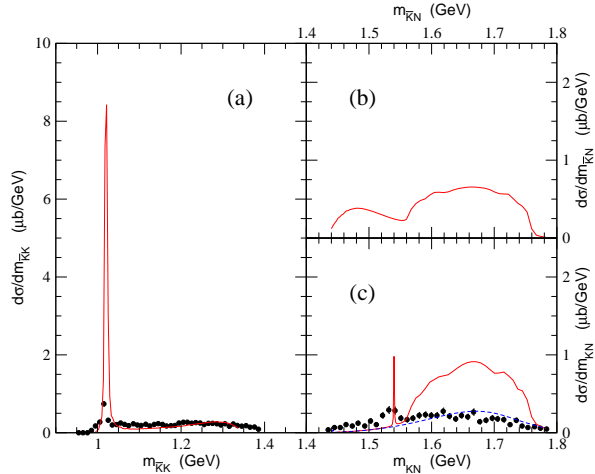, width=0.45\hsize, angle=-90}
\caption{
(a) $K\overline{K}$, (b) $\overline{K}N$, and (c) $KN$ invariant
mass distributions for $\gamma n \to K^+ K^- n$ at $E_\gamma = 2.3$ GeV.
The experimental data are from Refs.~\cite{CLAS03-b}. The dashed line in
(c) is obtained without the $\phi$ meson background and the $\Theta$
contributions. Here we assume that the $\Theta(1540)$ has even parity.}
\label{fig:2.3-3-even}
\end{figure}

\begin{figure}[t]
\centering
\epsfig{file=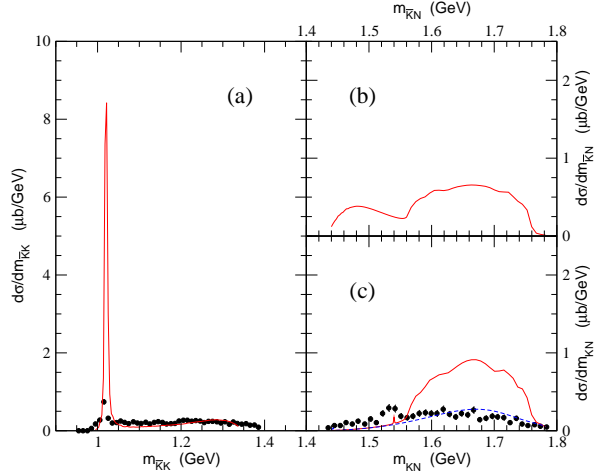, width=0.45\hsize, angle=-90}
\caption{
(a) $K\overline{K}$, (b) $\overline{K}N$, and (c) $KN$ invariant
mass distributions for $\gamma n \to K^+ K^- n$ at $E_\gamma = 2.3$ GeV.
The experimental data are from Refs.~\cite{CLAS03-b}.
Notations are the same as in Fig.~\ref{fig:2.3-3-even}.
Here we assume that the $\Theta(1540)$ has odd parity.}
\label{fig:2.3-3-odd}
\end{figure}

The results from assuming an even-parity (odd-parity) $\Theta(1540)$
are presented in Fig.~\ref{fig:2.3-3-even} (Fig.~\ref{fig:2.3-3-odd})
at $E_\gamma = 2.3$ GeV.
In both cases, we can reproduce very well the $\phi$ peak of the
$K^+K^-$ mass distributions [Fig.~\ref{fig:2.3-3-even}(a)
and Fig.~\ref{fig:2.3-3-odd}(a)].
The predicted shape in other region of the ${K^+K^-}$ invariant mass also
qualitatively agrees with the data.

The comparisons with the data of the $K^+n$ mass distributions are
shown in Fig.~\ref{fig:2.3-3-even}(c) and  Fig.~\ref{fig:2.3-3-odd}(c).
Note that the JLab data were obtained from removing the
$\phi\to K^+K^-$ decay contributions at the $\phi$ peak and hence should
only be compared with
the dashed curves which are obtained from turning off the $\phi$ and
$\Theta$ production contribution in our calculations.%
\footnote{We turn off the $\Theta$ contribution in the dashed curves in
Fig.~\ref{fig:2.3-3-even}(c) and  Fig.~\ref{fig:2.3-3-odd}(c) in order
to show the enhancement of the $\Theta$ peak compared with the backgrounds.}
For completeness, our full predictions (solid curves) are also displayed in
(c) of Fig.~\ref{fig:2.3-3-even} and Fig.~\ref{fig:2.3-3-odd}.
We see that the peak in $KN$ mass distribution arising from
the production of the $\Theta(1540)$ is not so much pronounced as in the
case of the $\phi$ meson peak in $K\overline{K}$ invariant mass
distribution.
This is mainly due to the small coupling of the $\Theta$ with $KN$ and $K^*N$.
We also observe in Fig.~\ref{fig:2.3-3-odd}(c) that the $\Theta(1540)$ peak is
much smaller in the case of odd-parity $\Theta$.
Shown in (b) of Figs.~\ref{fig:2.3-3-even} and \ref{fig:2.3-3-odd} are
the $K^-n$ mass distribution.

By comparing the dashed curves and the experimental data in
Fig.~\ref{fig:2.3-3-even}(c) and Fig.~\ref{fig:2.3-3-odd}(c),
we can see that the shape of the data can be reproduced well by our
model except in the region near the $\Theta(1540)$ peak. 
The width of the peak cannot be simply judged since the broad width of
the JLab data reflects the detector resolution.
Besides, there are two possible interpretations of our results.
First, the $\Theta(1540)$ is produced and the discrepancy between the data
and the obtained curves is due to the low statistics and limited
resolution of the experiment.
On the other hand, the discrepancy is perhaps due to the deficiency of
our model in accounting for other possible non-$\Theta$ mechanisms and
the existence of $\Theta$ is questionable, which also includes the
possibility of the contamination due to $\Lambda(1520)$.
As discussed in the previous Section, we are limited by the lack of
information in calculating some allowed non-$\Theta$ background
mechanisms.
Obviously, high statistics and high resolution experiments are
strongly required.
The existence of $\Theta(1540)$ can be unambiguously established if and
only if a very sharp resonance peak, which is very unlikely due to the
background amplitudes as predicted by our model, is observed.
Furthermore, an even-parity $\Theta$ is more likely to be detected, while
it will be difficult to identify an odd-parity $\Theta$, even if it exists,
from the background continuum.
This is because the odd-parity $\Theta$ has much smaller couplings with
the $KN$ channel than the even-parity $\Theta$ when the same decay width
is assigned.
Therefore, if the observed $\Theta(1540)$ has odd-parity, it must have
larger decay width, at least, one order of magnitude larger than the current
estimate in Ref.~\cite{PDG04} or it should have a large coupling to the
other channels~\cite{TEHN04}.

\begin{figure}[t]
\centering
\epsfig{file=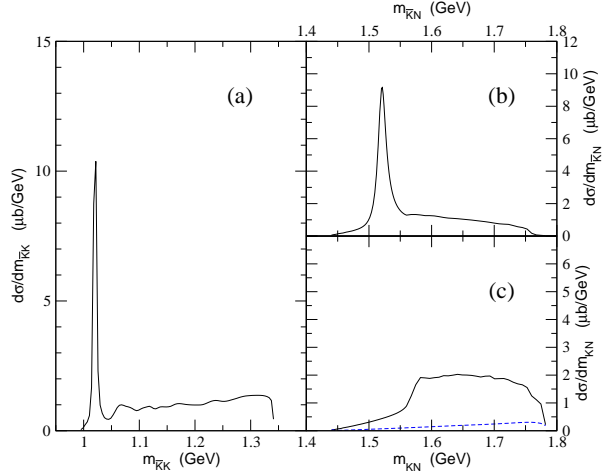, width=0.45\hsize, angle=-90}
\caption{
(a) $K\overline{K}$, (b) $\overline{K}N$, and (c) $KN$ invariant
mass distributions for $\gamma p \to K^+ K^- p$ at $E_\gamma = 2.3$ GeV.
The dashed line in (c) is obtained without the $\phi$ meson background and
the $\Lambda(1520)$ contribution.}
\label{fig:2.3-1}
\end{figure}

To facilitate the future experimental searches, we now present our predictions
for the other three processes listed in Eq.~(\ref{eq:channel}).
In Fig.~\ref{fig:2.3-1}, we give the results for the $\gamma p \to K^+ K^- p$
reaction at $E_\gamma = 2.3$ GeV.
Since we assume that $\Theta(1540)$ is isoscalar and can not contribute
to this reaction, there is no peak in $K^+p$ mass distribution.
The peak in Fig.~\ref{fig:2.3-1}(b) of the $K^-p$ mass distribution
is due to the $\Lambda(1520)$ of hyperon background (Fig.~\ref{fig:hyp}).
The dashed line in Fig.~\ref{fig:2.3-1}(c) is obtained when the contributions
from the $\phi$ and $\Lambda(1520)$ are neglected in the calculation.
Clearly, these two mechanisms are the major background production processes.
Experimental test of our predictions presented in Fig.~\ref{fig:2.3-1} will
also be an important task to check our model of non-$\Theta$ background
mechanisms which must be understood before the predictions for
$\gamma n \to K^+ K^- n$ can be used to determine the existence of the
$\Theta(1540)$.

\begin{figure}[t]
\centering
\epsfig{file=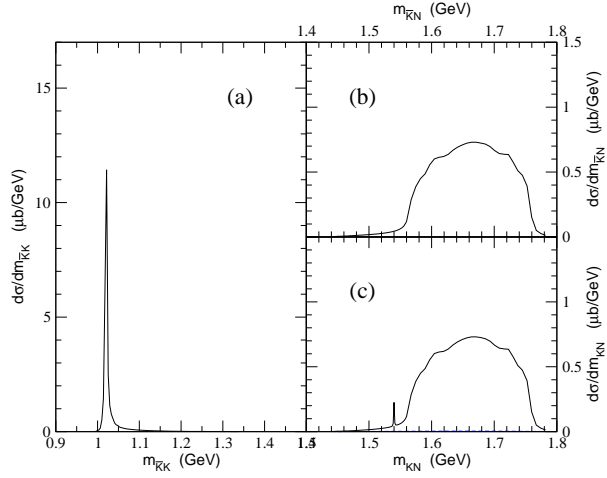, width=0.45\hsize, angle=-90}
\caption{(a) $K\overline{K}$, (b) $\overline{K}N$, and (c) $KN$ invariant
mass distributions for $\gamma p \to K^0 \overline{K}^0 p$ at
$E_\gamma = 2.3$ GeV.
The dashed line in (c) is obtained without the $\phi$ meson background and
the $\Theta(1540)$ contribution and magnified by a factor of 10.
Here we assume that the $\Theta(1540)$ has even parity.}
\label{fig:2.3-2-even}
\end{figure}

\begin{figure}[t]
\centering
\epsfig{file=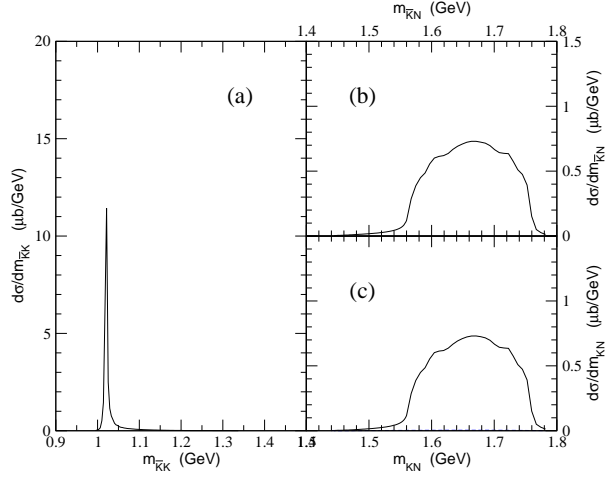, width=0.45\hsize, angle=-90}
\caption{(a) $K\overline{K}$, (b) $\overline{K}N$, and (c) $KN$ invariant
mass distributions for $\gamma p \to K^0 \overline{K}^0 p$ at
$E_\gamma = 2.3$ GeV.
The dashed line in (c) is obtained without the $\phi$ meson background and
the $\Theta(1540)$ contribution and magnified by a factor of 10.
Here we assume that the $\Theta(1540)$ has odd parity.
Since the contribution from $\Theta(1540)$ is suppressed, its peak is
not seen in (c).}
\label{fig:2.3-2-odd}
\end{figure}

The results for $\gamma p \to K^0\overline{K}^0 p$ are shown in
Figs.~\ref{fig:2.3-2-even} and \ref{fig:2.3-2-odd} for the even and odd
parity $\Theta$, respectively.
We see that the $\Theta(1540)$ peak in Fig.~\ref{fig:2.3-2-even} is much
smaller than that in the $\gamma n \to K^+ K^- n$ reaction
(Fig.~\ref{fig:2.3-3-even}).
The peak from the odd-parity $\Theta(1540)$ is hardly to be
seen in the $KN$ invariant mass distribution in $\gamma p \to K^0
\overline{K}^0 p$ as shown in Fig.~\ref{fig:2.3-2-odd}.

\begin{figure}[t]
\centering
\epsfig{file=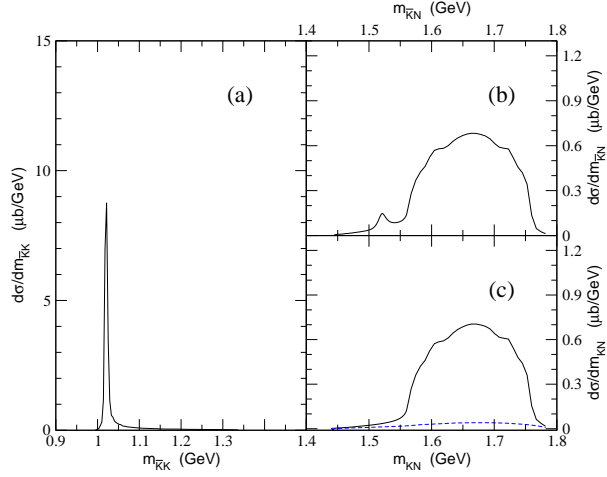, width=0.45\hsize, angle=-90}
\caption{(a) $K\overline{K}$, (b) $\overline{K}N$, and (c) $KN$ invariant
mass distributions for $\gamma n \to K^0 \overline{K}^0 n$ at
$E_\gamma = 2.3$ GeV.
The dashed line in (c) is obtained without the $\phi$ meson background
and the $\Lambda(1520)$ contribution.}
\label{fig:2.3-4}
\end{figure}

The results for $\gamma n \to K^0 \overline{K}^0 n$ are given in
Fig.~\ref{fig:2.3-4}.
Again, there is no contribution from the isoscalar $\Theta$ to this
reaction and therefore there is no peak in the $KN$ invariant mass
distribution [Fig.~\ref{fig:2.3-4}(c)].
Here we also find that the predicted cross sections [dashed curve in
Fig.~\ref{fig:2.3-4}(c)] is greatly reduced if the contributions from
the $\phi$ and $\Lambda(1520)$ productions are turned off.

\subsection{Tensor meson photoproduction contributions}

In this subsection, we discuss in more detail the contribution of the
tensor meson production to the $\gamma n \to K^+K^-n$ reaction, which is
an important issue raised by Dzierba {\it et al.\/}~\cite{DKST03},
who indicated the possibility that the observed peak at 1540 MeV
in the $KN$ invariant mass distribution could be a false peak arising from
tensor meson background.
Their calculation of neutral tensor meson photoproduction is based on 
a model of pion trajectory exchange mechanism.
However, as we discussed in Section~II, the pion exchange is not allowed for
this process because of the $C$ parity, and the lowest allowed exchanged
particles are vector mesons.
Although the production mechanism used by the authors of Ref.~\cite{DKST03}
is questionable, their claim should be checked by a calculation using the 
vector meson exchange mechanism, as formulated in Section II.2.B.

\begin{figure}[t]
\centering
\epsfig{file=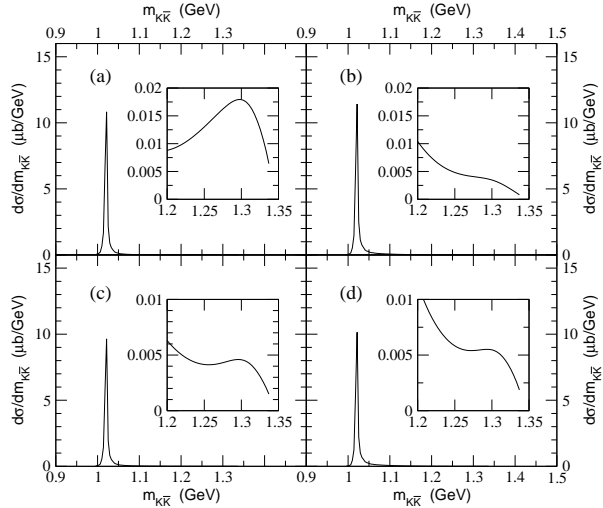, width=0.45\hsize, angle=-90}
\caption{Vector meson and tensor meson contributions to $K\overline{K}$
invariant mass distributions for (a) $\gamma p \to K^+ K^- p$,
(b) $\gamma p \to K^0 \overline{K}^0 p$ ,
(c) $\gamma n \to K^+ K^- n$,
(d) $\gamma n \to K^0 \overline{K}^0 n$ at $E_\gamma = 2.3$ GeV.
Shown in the small windows are the region of tensor meson peaks.}
\label{fig:KK-VT}
\end{figure}

Our calculations of the vector meson and tensor meson contributions to the
$K\overline{K}$ mass distributions of the four processes listed in
Eq.~(\ref{eq:channel}) are displayed in Fig.~\ref{fig:KK-VT}.
The contributions from tensor mesons (displayed in small windows of
Fig.~\ref{fig:KK-VT}) are clearly much smaller than the vector meson
contributions.
Furthermore, because of their large decay widths,
$\Gamma\bm{(}f_2(1275)\bm{)} \approx 185$ MeV and
$\Gamma\bm{(}a_2(1320)\bm{)} \approx 107$ MeV, the contributions from
these two tensor mesons do not give distinguishable two peaks in
$K\overline{K}$ mass distribution.
We also note that the tensor meson peak in $\gamma p \to K^+ K^- p$
reaction is much more pronounced than in the other reactions.
This is due to the isospin factors associated with the coupling constants,
which define the relative phases between different contributions in each
process.
As seen in Table~\ref{tab:TM}, the resulting relative phases lead to
constructive interference in $\gamma p \to K^+ K^- p$ reaction and
destructive interference in the other reactions.
As a result, the tensor meson peaks in the $\gamma n \to K^+ K^- n$
reaction is smaller than those in the $\gamma p \to K^+ K^- p$ reaction.
Therefore, the claim raised by Ref.~\cite{DKST03} can be checked by
comparing the results from the above two reactions.
Namely, if the peak at 1540 MeV in the $\gamma n \to K^+ K^- n$ reaction
is coming from the tensor meson contribution, one could expect a similar
or even more apparent peak at around 1540 MeV in $\gamma p \to K^+ K^- p$
reaction with the similar energy of the photon beam.
The absence of such a peak in $\gamma p \to K^+ K^- p$ reported by the
SAPHIR~\cite{SAPHIR03} and HERMES~\cite{HERMES-04} Collaborations,
therefore, seems to disfavor the possibility of ascribing the peak 
at 1540 MeV in the $KN$ mass distribution of $\gamma n \to K^+ K^- n$
to the tensor meson background.
This, of course, should be further examined by other higher statistics
experiments.

\begin{table}[t]
\centering
\begin{tabular}{c|ccc|ccc} \hline\hline
Reaction & \multicolumn{3}{c|}{$a_2^0$ photoproduction}
 & \multicolumn{3}{c}{$f_2^0$ photoproduction } \\ 
   & $\omega$ exchange & $\rho$ exchange & $a_2^0 \to K\overline{K}$ &
   $\omega$ exchange & $\rho$ exchange & $f_2^0 \to K\overline{K}$ 
\\ \hline
$\gamma p \to K^+ K^- p$ & $+$ & $+$ & $+$ & $+$ & $+$ & $+$ \\
$\gamma p \to K^0 \overline{K}^0 p$ & $+$ & $+$ & $-$ & $+$ & $+$ & $+$ \\
$\gamma n \to K^+ K^- n$ & $+$ & $-$ & $+$ & $+$ & $-$ & $+$ \\
$\gamma n \to K^0 \overline{K}^0 n$ & $+$ & $-$ & $-$ & $+$ & $-$ & $+$ \\
\hline\hline
\end{tabular}
\caption{Relative phases of tensor meson photoproduction contribution to
the $\gamma N \to K\overline{K}N$ reaction.}
\label{tab:TM}
\end{table}

\begin{figure}[t]
\centering
\epsfig{file=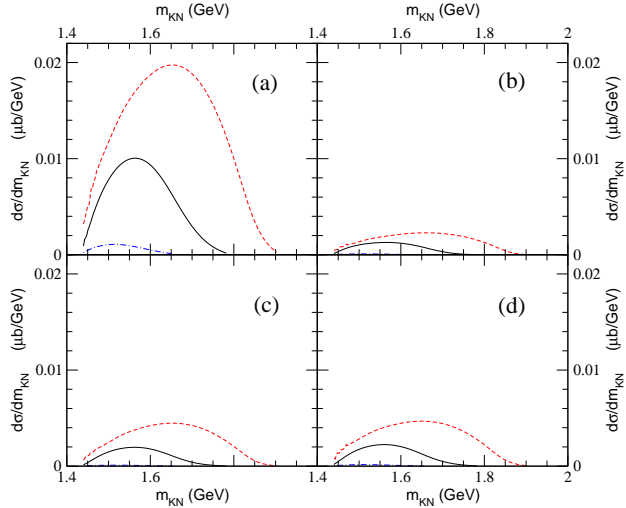, width=0.45\hsize, angle=-90}
\caption{
Tensor meson contribution to $KN$ invariant
mass distributions for (a) $\gamma p \to K^+ K^- p$,
(b) $\gamma p \to K^0 \overline{K}^0 p$ ,
(c) $\gamma n \to K^+ K^- n$,
(d) $\gamma n \to K^0 \overline{K}^0 n$ at $E_\gamma = 2.3$ GeV.
The dot-dashed, solid, and dashed lines are at $E_\gamma = 2.0$, $2.3$,
and $2.6$ GeV, respectively. The dot-dashed lines in (b,c,d) are
suppressed and hard to be seen within the given scale.}
\label{fig:KN-TM}
\end{figure}

In Fig.~\ref{fig:KN-TM}, we give our results for $KN$ invariant mass
distribution coming solely from the tensor meson photoproduction part.
Although their maximal values locate at around 1.56 GeV at $E_\gamma =
2.3$ GeV (solid curve), the shapes are very broad and most of 
the magnitudes are much smaller than the other backgrounds by about two
orders of magnitude.
Thus, the tensor meson contributions estimated within our model based
on vector meson exchange are too weak to generate any narrow peak in
the presence of other much larger background processes even if the
$\phi$ meson background is removed. 
We therefore conclude that we could not verify the claim of Ref.~\cite{DKST03}.
Our results also show that, if the peak is from the tensor meson background,
its position should change with different photon beam energies.

\subsection{Double differential cross sections}

The double differential cross sections are calculated for the three cases;
no pentaquark, even-parity $\Theta(1540)$, and odd-parity $\Theta(1540)$.
Since we are interested in the existence and parity of isoscalar $\Theta$,
we now focus on the $\gamma n \to K^+ K^- n$ reaction, which has larger
cross sections than $\gamma p \to K^0 \overline{K}^0 p$.

\begin{figure}[t]
\centering
\epsfig{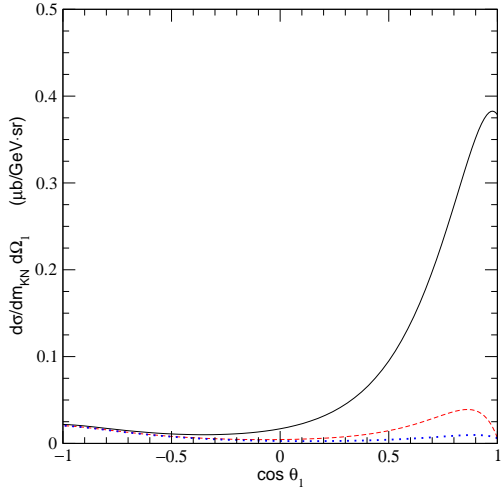}
\caption{
Double differential cross section $d\sigma/dm_{KN}d\Omega_1$ as
a function of $\cos\theta_1$, where $\theta_1$ is the polar angle of
$K^-$ in the center of mass frame, for $\gamma n \to K^+K^-n$ with
$E_\gamma = 2.3$ GeV and $m_{KN} = 1.54$ GeV.
The dotted line is obtained without $\Theta(1540)$, while the solid 
(dashed) line is with even (odd) parity of $\Theta$.}
\label{fig:diff2-2.3}
\end{figure}

Shown in Fig.~\ref{fig:diff2-2.3} is the double differential cross sections
for the $\gamma n \to K^+ K^- n$ reaction at $E_\gamma = 2.3$ GeV and
$m_{KN} = 1.54$ GeV, i.e., at the resonance point, as a function
of $\cos\theta_1$, where $\theta_1$ is the polar angle of $K^-$ in
the $\gamma N$ center of mass frame of which $z$ axis is defined as the
photon beam direction.
In this figure, the solid line is for even-parity $\Theta$ and the
dashed line for odd-parity $\Theta$. 
The dotted line is for the background, i.e., without $\Theta$.
Our result shows that the differential cross sections are enhanced by
the presence of even-parity $\Theta$ in the forward scattering region of
$K^-$ meson direction.
This suggests that the kinematic cut for the angle of $K^-$ would be useful
to enhance the $\Theta$ peak in $KN$ mass distributions.
But for the odd-parity $\Theta$, this enhancement is small.
This difference is primarily due to the magnitude of the $\Theta$
couplings.

We have also calculated some spin asymmetries of this reaction.
As discussed in Refs.~\cite{NL04,THH03,HBEK03}, however, we cannot avoid
model-dependence on the spin asymmetries in most reactions, which makes
it very hard to study the parity of $\Theta$ by the spin asymmetries
except some special cases like the $NN$ reaction near the threshold.
Therefore, we do not present our results on the spin asymmetries and do
not make a definite conclusion on the dependence of spin
asymmetries on the parity of $\Theta$.
Instead, we make a comment on the spin asymmetries of the model
considered in this work.
We have considered two spin asymmetries, the single photon beam asymmetry
$\Sigma_x$ and beam-target double asymmetry $C_{\rm BT}$ for the $\gamma
n \to K^+ K^- n$ reaction at the same energies as in Fig.~\ref{fig:diff2-2.3}.
(For their definitions, see Ref.~\cite{NT03}.)

The considered asymmetries, the photon beam asymmetry and beam-target
double asymmetry, were found to be rather sensitive to the parity of the
$\Theta(1540)$.
But in the case of beam-target double asymmetry, the difference between
the odd-parity $\Theta$ and the background processes is found to be not large.
Although the model-dependence is unavoidable, our model calculation
shows that these spin asymmetries may have different values
depending on the parity of $\Theta(1540)$, especially in the forward
scattering region.%
\footnote{
Here we note that our result on the photon beam asymmetry
agrees qualitatively with that of Ref.~\cite{NT03} which is based on
a much simpler model.
Although the numerical values and the structure at large $K^-$ angles are
different, these two models are consistent at least qualitatively at
forward angles of the $K^-$ momentum in the center of mass frame.
In Ref.~\cite{TEHN04}, it is claimed that the single and double spin
asymmetries are not sensitive to the parity of the $\Theta(1540)$.
We found that this comes mainly from the role of the $K^*$ exchanges,
Fig.~\ref{fig:theta}(f,g).
In Ref.~\cite{TEHN04}, the authors used the signal to background ratio
to constrain the ratio $\alpha \equiv g_{K^*N\Theta}/g_{KN\Theta}$, which
leads to $\alpha = 1.875$ and $8.625$ for the even and odd parity
$\Theta$. 
As a consequence, they have the $K^*$-exchange dominance in the $\Theta$
production mechanism especially in the case of odd-parity $\Theta$,
and as a result the single and double asymmetries are not sensitive to the
parity of $\Theta$.
In our calculation we used $\alpha = \sqrt3$ and $1/\sqrt3$ for the even
and odd parity $\Theta$ following the quark model predictions.
We also found that the spin asymmetries are sensitive to the parity of
$\Theta$ is $\alpha < 1$ for the odd parity $\Theta$.
Therefore, this illustrates the model-dependence of the asymmetries and it
is very important to estimate the correct order of magnitude of $\alpha$,
which is, however, not possible at the present state.
}

We have also calculated these asymmetries at a given angle $\theta_1$ as
functions of $m_{KN}$.
($\theta_1 = 30^\circ$ for $\Sigma_x$ and $\theta_1=0^\circ$ for $C_{\rm
BT}$)
We could find the resonance structure in these asymmetries when the
$\Theta$ has even parity.
For odd-parity $\Theta$, as can be inferred from Fig.~\ref{fig:diff2-2.3},
its contribution is small and the resonance structure in the beam-target
double asymmetry is not manifested.
Thus, although the parity of $\Theta(1540)$ may not be uniquely
determined by these asymmetries because of model-dependence \cite{NL04},
our results suggest that the resonance structure in the
asymmetries can be used to verify the existence of $\Theta$.

\section{Summary}

In this work, we have investigated how the $\gamma N \to K \overline{K} N$
reaction can be used to study the production and the properties of the
pentaquark $\Theta(1540)$ baryon.
We study the case that $\Theta(1540)$ is a spin $J=1/2$ and isospin $I=0$
particle, and have focused on the influence of the (non-$\Theta$) background
amplitudes due to the (tree) $t$-channel Drell diagrams
(Fig.~\ref{fig:Drell}), the $K\overline{K}$ production through the
intermediate vector meson and tensor meson photoproduction
(Fig.~\ref{fig:VTM}), and the mechanisms involving intermediate
$\Lambda(1116)$, $\Lambda(1405)$, $\Lambda(1520)$, $\Sigma (1193)$,
$\Sigma(1385)$, and $\Delta (1232)$ states (Fig.~\ref{fig:hyp}).
The vector meson photoproduction amplitude is calculated from a
phenomenological model which describes well the experimental data at low
energies.
The charged tensor meson production amplitude is calculated from a
one-pion-exchange model, which describes well the total cross section data of
$\gamma p \rightarrow a^{+}_2(1320) n$.
The neutral tensor meson production part is estimated by using the vector
meson exchange mechanisms.
The coupling constants needed for calculating all of the considered
background mechanisms are deduced from the data of decay widths by imposing
the SU(3) symmetry or making use of various quark model predictions.
No attempt is made to calculate other background amplitudes which are
kinematically allowed but can not be computed because of the lack of
experimental information.
Thus the present work represents only a step toward a complete
dynamical description of the $\gamma N \to K \overline{K} N$ reaction.
Nevertheless, some progress has been made in assessing the existing data
concerning the existence of the $\Theta(1540)$ pentaquark state.

The $\Theta(1540)$ production mechanism (Fig.~\ref{fig:theta}) is
calculated by taking 1 MeV as its decay width, which is close to the
Particle Data Group value.
With the background amplitude constrained by the total cross section data
of $\gamma p \to K^+K^-p$ (the considered isoscalar $\Theta(1540)$ is not
allowed in the process), we find that the resulting $K^+K^-$ and
$K^+ n$ invariant mass distributions of the $\gamma n \to K^+ K^- n$
reaction can qualitatively reproduce the shapes of the JLab data
although its magnitude cannot be compared.
However, the predicted $\Theta(1540)$ peak can not be identified
unambiguously with the data.
There are two possible interpretations of our results.
First, it is possible that the $\Theta(1540)$ is produced and the
discrepancy between the data and our results is due to the low statistics
and limited resolution of the experiment.
On the other hand, the discrepancy is perhaps due to the deficiency of
our model in accounting for other possible non-$\Theta$ mechanisms and
the existence of $\Theta(1540)$ is questionable.
Obviously, high statistics and high resolution experiments are needed.
The existence of $\Theta(1540)$ can be unambiguously established if and
only if a very sharp resonance peak, which is very unlikely from the
background amplitudes as predicted by our model, is observed.
We also find that an even-parity $\Theta$ is more likely to be detected,
while it will be difficult to identify an odd-parity $\Theta$, even if it
exists, from the background continuum, unless it has much larger coupling
to the $K^* N$ channel than the current quark model predictions and/or
other relevant production mechanism(s) not considered in this work.

We have analyzed in some detail the contributions of the tensor meson to
$\gamma n \rightarrow K^+K^-n$ reaction, which is an important issue raised
by Dzierba {\it et al.\/}~\cite{DKST03}. (See also Ref.~\cite{HBKS04}.)
These authors indicated that the observed peak at 1540 MeV in the $KN$
invariant mass distribution could be a false peak arising from tensor meson
background.
What we found are
(1) the calculation of Ref.~\cite{DKST03} is based on the $\pi^0$ exchange
mechanism which can not take place in neutral tensor meson photoproduction
if the $C$-parity is conserved.
(2) Instead of $\pi^0$ exchange, we estimate the neutral tensor meson 
contributions by using the vector meson exchange model, and have found that
the neutral tensor meson contribution is too weak to generate any
resonance peak which can be identified with the existing data of
$\gamma n \to K^+K^- n$ reaction.

Finally, we also present the double differential cross sections and make
a comment on the spin asymmetries.
Our results show that some kinematic cut would enhance the peak of
$\Theta$, if exists.
In the spin asymmetries, the major uncertainty comes from the ratio
$\alpha \equiv g_{K^*N\Theta}/g_{KN\Theta}$, and our results suggest a
resonance structure in some spin asymmetries, which can be measured at
current experimental facilities.

\acknowledgments

We are grateful to K. Hicks, K. Joo, T. Nakano, and
A.I. Titov for useful information and fruitful discussions.
We also thank V. Burkert for valuable comments.
Y.O. acknowledges fruitful discussions with Hungchong Kim.
This work was supported by Forschungszentrum-J{\"u}lich, contract No.
41445282 (COSY-058) and
U.S. DOE Nuclear Physics Division Contract No. W-31-109-ENG-38.

\appendix

\section{Tensor mesons}

In this Appendix, we discuss the couplings of tensor mesons to hadrons
and their radiative decays.
Among the spin-2 tensor mesons, we are interested in $f_2(1270)$,
$f_2'(1525)$, and $a_2(1320)$ as they can decay into two kaons and
close to the threshold energy region of the $\gamma N \to K\overline{K}N$
reaction.%
\footnote{The $f_2'(1525)$, which mostly decays into $K\overline{K}$, is
expected to be in $s\overline{s}$ state. Therefore, its coupling to the
nucleon would be suppressed. Because of its higher mass and assuming the
OZI rule as the first approximation, we do not include $f_2'$
photoproduction in our study of $K\overline{K}$ photoproduction
processes.}
The $f_2(1270)$ and $f_2'(1525)$ have the quantum numbers $I^G(J^{PC})
= 0^+(2^{++})$.
The $f_2(1270)$ has mass $M_{f_2} = 1275$ MeV, width $\Gamma_{f_2} =
185.1$ MeV, and it mostly decays into two pions.
Some of its branching ratios are
$\mbox{BR}(f_2 \to \pi\pi) = (84.7 \stackrel{+2.4}{\mbox{\scriptsize
$-1.3$}})\%$,
$\mbox{BR}(f_2 \to K\overline{K}) = (4.6 \pm 0.5)\%$,
and
$\mbox{BR}(f_2 \to \gamma\gamma) = (1.41 \pm 0.13) \times 10^{-5}$.

The $f_2'(1525)$ has $M_{f_2'} = 1525$ MeV and $\Gamma_{f_2'} = 76 \pm
10$ MeV.
It mostly decays into two kaons and some of its branching ratios are
$\mbox{BR}(f_2' \to K\overline{K}) = (88.8 \pm 3.1)\%$ and
$\mbox{BR}(f_2' \to \gamma\gamma) = (1.23 \pm 0.17) \times 10^{-6}$.

The $a_2^{}(1320)$ is an isovector tensor meson with $M_{a_2^{}} = 1318$ MeV
and $\Gamma_{a_2^{}} = 107 \pm 5$ MeV.
It mostly decays into $\rho\pi$, but it decays also into two kaons,
$\pi\gamma$, and two photons with
$\mbox{BR}(a_2^{} \to K\overline{K}) = (4.9 \pm 0.8)\%$,
$\mbox{BR}(a_2^{} \to \pi^\pm\gamma) = (2.68 \pm 0.31) \times 10^{-3}$,
and
$\mbox{BR}(a_2^{} \to \gamma\gamma) = (9.4 \pm 0.7) \times 10^{-6}$.
Note that $a_2^0 \to \pi^0 \gamma$ decay is not allowed because of
$C$-parity.
By the same reason, $f_2 \to \pi^0 \gamma$ is forbidden.

In our calculation, we use the coupling constants determined from
the experimental data and quark model predictions.
Here, we discuss the way to determine the couplings and compare the
values with those obtained by assuming SU(3) symmetry and vector meson
dominance.

\subsection{Interactions with pseudoscalar mesons}

Since $f_2$ and $a_2$ are spin-2 tensor mesons, we are dealing with
tensor meson nonet whose members are $a_2(1320)$, $K_2(1430)$,
$f_2(1275)$, and $f_2'(1525)$.
This is analogous to vector meson nonet of $\rho$, $K^*$, $\omega$, and
$\phi$.

The pseudoscalar octet is represented by an SU(3) matrix $P$ which is
defined in Eq.~(\ref{eq:octet}).
Similarly, the tensor meson octet is represented by $T_8$ as
\begin{equation}
T_8 = \left( \begin{array}{cc} 
 \frac{1}{\sqrt6} f_8 + \frac{1}{\sqrt2} a_2 & K_2 \\
\overline{K}_2 & - \frac{2}{\sqrt6} f_8
\end{array} \right),
\end{equation}
where $a_2 = \bm{a}_2 \cdot \bm{\tau}$.
The Lorentz index $\mu\nu$ is suppressed.
Then the $T_8PP$ interaction is obtained as 
\begin{equation}
\mathcal{L} = g \mbox{Tr}\, \left( T_8^{\mu\nu} \partial_\mu P
\partial_\nu P \right),
\end{equation}
which gives the SU(3) symmetry relations to the coupling constants.

Since $f_2(1275)$ and $f_2'(1525)$ are expected to be close to ideal
mixing, we introduce the tensor meson singlet $f_0$, whose interaction
with two pseudoscalar mesons is given by
\begin{eqnarray}
\mathcal{L}_{f_0PP} &=& \frac{g}{\sqrt3} \mbox{Tr} \left( f_0^{\mu\nu}
\partial_\mu P \partial_\nu P \right)
\nonumber \\
&=& \frac{g}{\sqrt3} f_0^{\mu\nu} \left\{ \partial_\mu \eta
\partial_\nu \eta + \partial_\mu \bm{\pi} \cdot \partial_\nu \bm{\pi} +
2 \partial_\mu \overline{K} \partial_\nu K \right\}.
\end{eqnarray}

Now we consider the mixing of $f_8$ and $f_0$.
The physical states $f_2$ and $f_2'$ are written as
\begin{eqnarray}
f' &=& \cos\theta f_8 - \sin\theta f_0, \nonumber \\
f &=& \sin\theta f_8 + \cos\theta f_0,
\end{eqnarray}
where $f_2$ and $f_2'$ are denoted by $f$ and $f'$, respectively.
Then we have
\begin{eqnarray}
&& g_{f'\pi\pi} = \frac{g}{\sqrt3} \left( \frac{1}{\sqrt2} \cos\theta -
\sin\theta \right), \qquad
g_{f\pi\pi} = \frac{g}{\sqrt3} \left( \frac{1}{\sqrt2} \sin\theta +
\cos\theta \right),
\nonumber \\
&& g_{f'KK} = -\frac{g}{\sqrt3} \left( \frac{1}{\sqrt2} \cos\theta +
2\sin\theta \right), \qquad
g_{fKK} = \frac{g}{\sqrt3} \left( -\frac{1}{\sqrt2} \sin\theta +
2\cos\theta \right).
\label{coup1}
\end{eqnarray}

The mixing angle can be estimated from the masses of tensor meson
nonet;
\begin{equation}
\tan^2\theta = \frac{3 M_{f'}^2 - 4 M_{K_2}^2 + M_{a_2}^2}{4 M_{K_2}^2
- M_{a_2}^2 - 3 M_f^2} \approx 0.35,
\end{equation}
where we have used the Gell-Mann--Okubo mass relation for squared
masses of tensor mesons.
This gives us
\begin{equation}
\theta \approx 30.5^\circ,
\end{equation}
where the ideal mixing angle is
$\theta_{\rm ideal mixing} \approx 35.3^\circ$.
Note that in the case of vector meson nonet, the mixing angle is
$\theta_V \approx 40^\circ$. Therefore, we can see that the tensor meson
nonet is as close to ideal mixing as the vector meson nonet.%
\footnote{Note that we are assuming $q\overline{q}$ structure of the
tensor meson, which may contain non-negligible glueball component.}

The $2^+0^-0^-$ interaction is obtained as
\begin{equation}
\mathcal{L} = - \frac{2G_{f\pi\pi}}{M_f} \partial_\mu \bm{\pi} \cdot
\partial_\nu \bm{\pi} f^{\mu\nu},
\end{equation}
which gives the decay width of $f_2 \to \pi\pi$ as
\begin{equation}
\Gamma(f_2 \to \pi\pi) = \frac{2}{5\pi}\frac{G_{f\pi\pi}^2}{M_f^4}
p_F^5,
\end{equation}
where
$p_F^{} = M_f \sqrt{ 1/4 - {M_\pi^2}/{M_f^2}}$.
Using $\Gamma(f_2\to \pi\pi)_{\rm expt.} \approx 156.9$ MeV and
$\Gamma(f_2' \to \pi\pi)_{\rm expt.} \approx 0.623$ MeV, we obtain
\begin{equation}
G_{f\pi\pi} = 5.76, \qquad
G_{f'\pi\pi} = 0.33 = 0.06 \ G_{f\pi\pi}.
\end{equation}
Similarly, $f \to K \overline{K}$ vertex can be obtained from
\begin{equation}
\mathcal{L} = - \frac{2G_{fKK}}{M_f} \partial_\mu \overline{K}
\partial_\nu K f^{\mu\nu}.
\end{equation}
Thus the decay width $f_2 \to K \overline{K}$ is estimated as
\begin{equation}
\Gamma(f_2 \to K\overline{K}) = \frac{2}{15\pi} \frac{G_{fKK}^2}{M_f^4}
p_F^5,
\end{equation}
which gives
\begin{equation}
G_{fKK} = 7.15 = 1.24 G_{f\pi\pi}.
\end{equation}
with $\Gamma(f_2 \to K\overline{K})_{\rm expt.} \approx 8.6 \pm 0.8$ MeV.
This expression can also be applied to $f_2'(1525)$ meson decay, and
with $\Gamma(f_2' \to K\overline{K})_{\rm expt.} \approx 65 \pm 5$ MeV 
we get
\begin{equation}
G_{f'KK} = -11.23 = -1.95 G_{f\pi\pi},
\end{equation} 
where the interaction is given by
\begin{equation}
\mathcal{L} = - \frac{2G_{f'KK}}{M_{f'}} \partial_\mu \overline{K}
\partial_\nu K f'^{\mu\nu}.
\end{equation}
Note that $f_2$ and $f_2'$ has the same quantum number and they are
anticipated to form ideal mixing as in the case for $\omega$ and $\phi$
mesons. This is because $f_2$ mostly decays into two pions, while
$f_2'$ mostly into two kaons. But the presence of $f_2 \to K\overline{K}$
decay implies the deviation from the ideal mixing.

In the case of $a_2(1320)$, because of its isovector nature, the
interaction Lagrangian reads
\begin{equation}
\mathcal{L} = - \frac{2G_{aKK}}{M_a} \partial_\mu \overline{K}
\bm{\tau} \cdot \bm{a}^{\mu\nu} \partial_\nu K.
\end{equation}
Using $\Gamma(a_2 \to K\overline{K}) \approx 5.24$ MeV, we obtain
\begin{equation}
G_{aKK} = 4.89 = 0.85 G_{f\pi\pi}.
\end{equation}

The obtained results should be compared with the SU(3) symmetry
relations, $G_{f'\pi\pi} = 0.10 \ G_{f\pi\pi}$,
$G_{fKK} = 1.11 \ G_{f\pi\pi}$, $G_{f'KK} = -1.60 \ G_{f\pi\pi}$, and
$G_{aKK} = 1.04 \ G_{f\pi\pi}$.
The deviation from the SU(3) symmetry relations implies the SU(3)
symmetry breaking effects and possibly the non-negligible glueball components
in tensor mesons.
Here, in the study of $K\overline{K}$ photoproduction, we use the coupling
constants determined from the measured decay widths of tensor mesons.

\subsection{Tensor meson radiative decays}

First, we consider the $2^+0^-1^-$ interaction such as $a_2 \to
\rho\pi$ or $a_2 \to \gamma\pi$ decays.
The interaction Lagrangian reads \cite{HHKS66}
\begin{equation}
\mathcal{L}_{a_2\gamma\pi} = \frac{g_{a_2\gamma\pi}}{M_a^2}
\varepsilon^{\mu\nu\alpha\beta} \partial_\mu A_\nu a^\pm_{\alpha\rho}
(\partial^\rho \partial_\beta \pi^\mp).
\end{equation}
The coupling constant $g_{a_2\gamma\pi}^{}$ is determined from the decay
width of $a_2^\pm \to \pi^\pm\gamma$ as
\begin{equation}
g_{a_2\gamma\pi} \approx 0.96.
\end{equation}

For the decays of a tensor meson into two photons, the most general form
reads \cite{Renn71}
\begin{equation}
\langle \gamma(k) \gamma(k') | T \rangle = \frac{1}{M_T} \epsilon^\kappa
\epsilon'^\lambda \epsilon^{\mu\nu} A_{\kappa\lambda\mu\nu}(k,k'),
\end{equation}
where the form of $A^{}_{\kappa\lambda\mu\nu} (k,k')$ is given by
Eq.~(\ref{eq:TVG}).
With the above interaction, we have
\begin{equation}
\Gamma(T \to \gamma\gamma) = \frac{M_f}{20\pi} \left( \frac{1}{24}
f_{T\gamma\gamma}^2 + g_{T\gamma\gamma}^2 \right).
\end{equation}
Since
\begin{eqnarray}
\Gamma(f_2 \to \gamma\gamma)_{\rm expt.} &=& 2.6 \pm 0.24 \mbox{ keV},
\nonumber \\
\Gamma(a_2 \to \gamma\gamma)_{\rm expt.} &=& 1.0 \pm 0.06 \mbox{ keV},
\nonumber \\
\Gamma(f_2' \to \gamma\gamma)_{\rm expt.} &=& 9.35 \times 10^{-2} \mbox{
keV},
\end{eqnarray}
we get
\begin{equation}
g_{f\gamma\gamma}^{} = 0.011, \qquad g_{a\gamma\gamma}^{} = 6.9 \times
10^{-3}, \qquad g_{f'\gamma\gamma} ^{}= -1.96 \times 10^{-3},
\end{equation}
assuming $f_{T\gamma\gamma}^{} = 0$.

The above interaction form can be used for the interactions of a tensor
meson with a vector meson and a photon, which gives
\begin{equation}
\Gamma(T \to V\gamma) = \frac{M_T^{}}{10\pi} g_{TV\gamma}^2 (1-x)^3
\left( 1 + \frac{x}{2} + \frac{x^2}{6} \right),
\end{equation}
with $x = M_V^2/M_T^2$.
There is no experimental data for this decay, so we use the predictions
of Ref.~\cite{IYO89} based on a covariant quark model, which gives a
reasonable description of the known radiative decay widths of vector
and tensor mesons.
The predictions and the obtained coupling constants are given in
Table~\ref{tab:TVG}.

\section{Effective Lagrangians for the hyperon backgrounds}

In this Appendix, we present the effective Lagrangians and coupling
constants used in the calculation for the hyperon background diagrams,
Fig.~\ref{fig:hyp}.

\subsection{Baryon octet}

The baryon octet included in our calculation of Figs.~\ref{fig:hyp}(a)-(e)
are the nucleon ($N$), $\Lambda(1116)$, and $\Sigma(1193)$. 
Their interactions with the photon are defined by
\begin{eqnarray}
\mathcal{L}_{\gamma NN} &=& -e \overline{N} \left[ A_\mu \gamma^\mu
\frac{1+\tau_3}{2} - \frac{1}{2M_N} (\kappa_s^N + \kappa_v^N \tau_3)
\sigma_{\mu\nu} \partial^\nu A^\mu \right] N,
\nonumber \\
\mathcal{L}_{\gamma\Lambda\Lambda} &=& \frac{e\kappa^\Lambda}{2M_N}
\overline{\Lambda} \sigma_{\mu\nu} \partial^\nu A^\mu \Lambda,
\nonumber \\
\mathcal{L}_{\gamma\Sigma\Sigma} &=& -e \overline{\Sigma}
\left[ A_\mu \gamma^\mu T_3 - \frac{1}{2M_N}
(\kappa_s^\Sigma + \kappa_v^\Sigma T_3)
\sigma_{\mu\nu} \partial^\nu A^\mu \right] \Sigma,
\end{eqnarray}
where $T^3 = \mbox{diag}(1,0,-1)$ and
\begin{eqnarray}
&& \kappa_s^N = \frac12(\kappa_p^{} + \kappa_n^{}) = -0.06, \qquad
\kappa_v^N = \frac12(\kappa_p^{} - \kappa_n^{}) = 1.85, \nonumber \\
&& \kappa^\Lambda = -0.61, \qquad
\kappa_s^\Sigma = \frac12(\kappa_{\Sigma^+}^{} + \kappa_{\Sigma^-}^{}) =
0.65, \qquad
\kappa_v^\Sigma = \frac12(\kappa_{\Sigma^+}^{} - \kappa_{\Sigma^-}^{}) = 0.81.
\end{eqnarray}
The numbers above are obtained by using the measured magnetic moments of
the baryon octet~\cite{PDG04}, $\mu(p) = 2.79$, $\mu(n) = -1.91$,
$\mu(\Lambda) = -0.61$, $\mu(\Sigma^+) = 2.46$, and $\mu(\Sigma^-) = -1.16$
in the nucleon magneton unit.
To calculate the photo-transition of $\Sigma$ into $\Lambda$ in
Fig.~\ref{fig:hyp}(d), we use 
\begin{equation}
\mathcal{L}_{\gamma\Sigma\Lambda} = \frac{e\mu_{\Sigma\Lambda}^{}}{2M_N}
\overline{\Sigma}^0 \sigma_{\mu\nu} \partial^\nu A^\mu \Lambda + \mbox{
H.c.},
\end{equation}
where $\mu_{\Sigma\Lambda}^{} = -1.61 \pm 0.08$ as given by the Particle
Data Group~\cite{PDG04}.

For meson-baryon interactions, we use the pseudoscalar coupling to write
\begin{eqnarray}
\mathcal{L}_{KN\Lambda} &=& -i g_{\Lambda NK}^{} \overline{N} \gamma_5
\Lambda K + \mbox{ H.c.},
\nonumber \\
\mathcal{L}_{KN\Sigma} &=& -i g_{\Sigma NK}^{} \overline{N} \gamma_5
\bm{\tau} \cdot \bm{\Sigma} K + \mbox{ H.c.}.
\end{eqnarray}
The flavor SU(3) symmetry relations evaluated with $d+f=1$ give
\begin{equation}
g_{KN\Lambda}^{} = -\frac{1}{\sqrt3} (1+2f) g_{\pi NN}^{}, \qquad
g_{KN\Sigma}^{} = (1-2f) g_{\pi NN}.
\end{equation}
By using the empirical value $f/d=0.575$,
 which gives $f=0.365$ and $d=0.635$, and $g_{\pi
NN}^2/4\pi = 14$, we get
\begin{equation}
g_{KN\Lambda}^{} = -13.24, \qquad g_{KN\Sigma}^{} = 3.58.
\end{equation}

To calculate Figs.~\ref{fig:hyp}(h) and (i) with intermediate $K^*$,
we use the following Lagrangian,
\begin{equation}
\mathcal{L}_{K^*NY} = -g_{K^*NY}^{} \overline{N} \left( \gamma_\mu Y
K^{*\mu} + \frac{\kappa^{}_{K^*NY}}{2M_N} \sigma_{\mu\nu} Y \partial^\nu
K^{*\mu} \right).
\end{equation}
We use the values from the new Nijmegen potential~\cite{RSY99,SR99}
to define the coupling constants in the above equation,
\begin{eqnarray}
g_{K^*N\Lambda}^{} = -6.11 \sim -4.26, &\qquad&
\kappa_{K^*N\Lambda}^{} = 0.436 \sim 0.474, \nonumber \\
g_{K^*N\Sigma}^{} = -3.52 \sim -2.46, &\qquad&
\kappa_{K^*N\Sigma }^{} = -1.0 \sim -0.412.
\end{eqnarray}
For our numerical calculation, the values in the right boundary are
used.

\subsection{Baryon decuplet}

We now consider the couplings involving 
the members of the baryon decuplet, $\Delta(1232)$ and
$\Sigma^*(1385)$, which are intermediate states in 
Figs.~\ref{fig:hyp}(a)-(g).
The Lagrangians describing the photo-interaction of $\Sigma^*(1385)$
read 
\begin{eqnarray}
\mathcal{L}_{\gamma \Sigma^* \Sigma^*} &=& e \overline{\Sigma}^*_\mu
A_\alpha \Gamma^{\alpha,\mu\nu}_{\gamma\Sigma^*} \Sigma^*_\nu,
\nonumber \\
\nonumber \\
\mathcal{L}_{\gamma KN\Sigma^*} &=& -i\frac{e f_{KN\Sigma^*}}{M_K} A^\mu
\left( \overline{\Sigma}_\mu^{*0} p K^- + \sqrt2
\overline{\Sigma}_\mu^{*-} n K^- - K^+ \overline{p} \Sigma_\mu^{*0} -
\sqrt2 K^+ \overline{n} \Sigma_\mu^{*-} \right),
\label{eq:Lag-RS}
\end{eqnarray}
where~\cite{LLD91}%
\footnote{There is an ambiguity with defining the non-minimal anomalous
magnetic moment term. Here we follow Ref.~\cite{LLD91}. Other
definitions are discussed in Refs.~\cite{ELP92,DWP00}.}
\begin{equation}
A_\alpha \Gamma^{\alpha,\mu\nu} = \left\{
g^{\mu\nu} \gamma^\alpha - \frac16 ( \gamma^\mu
\gamma^\nu \gamma^\alpha + \gamma^\alpha \gamma^\mu \gamma^\nu) \right\}
A_\alpha T_3
- \frac{1}{2M_N} \left( \kappa_s^{\Sigma^*} + \kappa_v^{\Sigma^*} T_3
  \right) \sigma^{\alpha\beta} \partial_\beta A_\alpha g^{\mu\nu},
\end{equation}
and the coupling $f_{KN\Sigma^*}$ will be explained later.
Since there is no experimental information for the magnetic moments of
$\Sigma(1385)$, we make use of the following quark model
predictions~\cite{Lic77}
\begin{equation}
\mu(\Sigma^{*+}) = 3.15, \qquad
\mu(\Sigma^{*0}) = 0.36, \qquad
\mu(\Sigma^{*-}) = -2.43,
\end{equation}
to obtain
\begin{equation}
\kappa_s^{\Sigma^*} = 0.36,
\qquad
\kappa_s^{\Sigma^*} = 1.79.
\end{equation}

We next need to construct the photo-transition Lagrangians
$\mathcal{L}_{\gamma N\Delta}$, $\mathcal{L}_{\gamma \Lambda\Sigma^*}$, and
$\mathcal{L}_{\gamma\Sigma\Sigma^*}$.
These can be fixed by considering the radiative decays of the decuplet baryons.
The most well-studied is the $\Delta \to N \gamma$ transition which
enters into Figs.~\ref{fig:hyp}(c) and (e).
Here we follow Ref.~\cite{DMW91} to write
\begin{eqnarray}
\mathcal{L}_{\gamma N \Delta}^1 &=& \frac{ie g_{1N\Delta}^{}}{2M_N}
\overline{\Delta}^\mu O_{\mu\lambda}(Z) \gamma_\nu \gamma_5
I^3(\textstyle\frac32,\frac12)
N F^{\nu\lambda} + \mbox{H.c.},
\nonumber \\
\mathcal{L}_{\gamma N \Delta}^2 &=& -\frac{e g_{2N\Delta}^{}}{4M_N^2}
\overline{\Delta}^\mu O_{\mu\nu}(Z) \gamma_5
I^3(\textstyle\frac32,\frac12)
(\partial_\lambda N) F^{\nu\lambda} + \mbox{H.c.},
\end{eqnarray}
where $F^{\mu\nu}$ is the field strength tensor of the photon,
$F^{\mu\nu} = \partial^\mu A^\nu - \partial^\nu A^\mu$, and we choose
the off-shell parameters so that $O_{\mu\nu}(Z) = g_{\mu\nu}$.
The isospin factor is calculated as
\begin{equation}
\overline{\Delta} I^3(\textstyle\frac32,\frac12) N = \sqrt{\frac23}
\left( \overline{\Delta}^+ p + \overline{\Delta}^0 n \right).
\end{equation}
The above Lagrangians lead to the following expression for calculating
the radiative decay width of the $\Delta$,
\begin{eqnarray}
\Gamma(\Delta \to N\gamma) &=& \frac{p_\gamma^3}{72\pi M_\Delta^2} \left(
\frac{e}{2M_N} \right)^2 \Biggl\{ \left[ g_{1N\Delta} (3 M_\Delta +
M_N) - g_{2N\Delta}^{} \frac{M_\Delta}{2M_N} (M_\Delta - M_N) \right]^2
\nonumber \\ && \mbox{} \qquad\qquad\qquad
+ 3 \left[ g_{1N\Delta}^{} - g_{2N\Delta}^{} \frac{M_\Delta}{2M_N}
\right]^2 (M_\Delta - M_N )^2 \Biggr\}.
\end{eqnarray}
It is well-known that the $g_{2N\Delta}^{}$ term is sensitive to the $E2/M1$
ratio of $\Delta \to N\gamma$ decay. In this calculation, however, we
assume that $g_{2N\Delta}^{} = 0$, which gives somewhat large value,
$\approx -6\%$, of $E2/M1$.
However, our results show that the contributions from the $\Delta N$
transition to $\gamma N \to K\overline{K} N$ reaction are suppressed
compared with the other contributions.
Therefore the precise value of $g_{2N\Delta}^{}$ is irrelevant to this
calculation.
Using $\Gamma(\Delta \to N\gamma)_{\rm expt.} \approx 672$ keV, we then get
\begin{equation}
g_{N\Delta} = g_{1N\Delta}^{} \approx 4.9,
\end{equation}
which is close to the fitted value, $\approx 5.0$, of Ref.~\cite{DMW91}.

To calculate $Y \to Y' \gamma$ transition in Fig.~\ref{fig:hyp}(d), we
consider the following Lagrangian,
\begin{equation}
\mathcal{L}_{\gamma BD} = \frac{ie g_{BD}^{}}{2M_N}
\overline{D}^\mu O_{\mu\lambda}(Z) \gamma_\nu \gamma_5
B F^{\nu\lambda} + \mbox{H.c.},
\end{equation}
where $D^\mu$ is the $\Sigma^*(1385)$ and $B$ denotes for
$\Lambda(1116)$ or $\Sigma(1193)$.
There is no experimental information on the radiative decay widths for
decuplet $\Sigma^*$(1385) except some upper bounds.
We therefore make use of the following quark model predictions~\cite{DHK83},
\begin{eqnarray}
&& \Gamma(\Sigma^{*0} \to \Lambda\gamma) = 232 \mbox{ keV}, \qquad
\Gamma(\Sigma^{*+} \to \Sigma^+\gamma) = 104 \mbox{ keV},
\nonumber \\ &&
\Gamma(\Sigma^{*0} \to \Sigma^0\gamma) = 19 \mbox{ keV}, \qquad
\Gamma(\Sigma^{*-} \to \Sigma^-\gamma) = 2.5 \mbox{ keV}.
\end{eqnarray}
The above values then fix the coupling constants of $\mathcal{L}_{\gamma BD}$
as
\begin{eqnarray}
g_{\Sigma^{*0}\Lambda} \approx 3.01, \qquad
g_{\Sigma^{*+}\Sigma^+} \approx 3.38, \qquad
g_{\Sigma^{*0}\Sigma^0} \approx 1.44, \qquad
g_{\Sigma^{*-}\Sigma^-} \approx 0.52.
\end{eqnarray}

We now will use the SU(3) symmetry to fix the couplings involving
$K$ mesons, baryon decuplet [$\Sigma^*(1385)$, $\Delta(1232)$] and
baryon octet [$N$, $\Lambda(1116)$, $\Sigma(1193)$].
We start with the well-studied Lagrangian for the $\pi N\Delta$
interaction.
We follow Refs.~\cite{BDM89,BMZ95} to write
\begin{equation}
\mathcal{L}_{\pi N\Delta} = \frac{f_{\pi N \Delta}}{m_\pi}
\overline{\Delta}^\mu O_{\mu\nu}(Z)
\bm{I}(\textstyle\frac32,\frac12) \cdot
\partial_\nu \bm{\pi} N + \mbox{H.c.},
\end{equation}
where the isospin transition matrix reads
\begin{equation}
\bm{I}(\textstyle\frac32,\frac12) \cdot \bm{\pi}
 = - I^{(+1)}_{3/2,1/2} \pi^+ +
I^{(-1)}_{3/2,1/2} \pi^- +
I^{(0)}_{3/2,1/2} \pi^0,
\label{t31}
\end{equation}
with
\begin{eqnarray}
I^{(+1)}_{\textstyle\frac32,\frac12} = \frac{1}{\sqrt6}
\left( \begin{array}{cc} \sqrt6 & 0 \\ 0 & \sqrt2 \\ 0 & 0 \\ 0 & 0
\end{array} \right), \quad
I^{(0)}_{\textstyle\frac32,\frac12} =
\frac{1}{\sqrt6} \left( \begin{array}{cc} 0 & 0 \\ 2 & 0 \\ 0 & 2
\\ 0 & 0 \end{array} \right), \quad
I^{(-1)}_{\textstyle\frac32,\frac12} =
\frac{1}{\sqrt6} \left( \begin{array}{cc} 0 & 0 \\ 0 & 0 \\ \sqrt2
& 0 \\ 0 & \sqrt6 \end{array} \right).
\end{eqnarray}
With $O_{\mu\nu}(Z) = g_{\mu\nu}$~\cite{BDM89,BMZ95}
(see also Ref.~\cite{ELP92}),
the $\Delta \to \pi N$ decay width can be written as
\begin{equation}
\Gamma(\Delta \to N\pi) = \frac{p_\pi^3}{24\pi} \left(
\frac{f_{\pi N \Delta}}{M_\pi} \right)^2 \frac{1}{M_\Delta^2} \left[
(M_\Delta + M_N)^2 - M_\pi^2 \right].
\end{equation}
Using $M_\Delta = 1232$ MeV and $\Gamma(\Delta \to N\pi) = 120$ MeV, we
get%
\footnote{If we use the pole mass $M_\Delta = 1211$ MeV, we get
$f_{\pi N \Delta}' = 2.56$.
In Ref. \cite{SL96}, $f_{\pi N\Delta}$ is estimated to be $2.05$--$2.12$.}
\begin{equation}
f_{\pi N\Delta} = 2.23.
\end{equation}
With the $\pi N \Delta$ coupling fixed, we then use the SU(3) relations,
\begin{equation}
\frac{f_{K\Sigma\Delta}}{M_K} = - \frac{f_{\pi N\Delta}}{M_N}, \qquad
f_{KN\Sigma^*} = \frac{1}{\sqrt6} f_{K\Sigma\Delta},
\end{equation}
to obtain
\begin{eqnarray}
\mathcal{L}_{K\Sigma\Delta} &=& \frac{f_{K\Sigma\Delta}}{M_K}
\overline{\Delta}^\mu \bm{I}(\textstyle\frac32,\frac12) \cdot
\bm{\Sigma} \partial_\mu K + \mbox{H.c.},
\nonumber \\
\mathcal{L}_{KN\Sigma^*} &=& \frac{f_{KN\Sigma^*}}{M_K} \partial_\mu
\overline{K} \, \overline{\bm{\Sigma}}^{*\mu} \cdot \bm{\tau} N + \mbox{ H.c.}.
\end{eqnarray}
where 
\begin{equation}
f_{K\Sigma\Delta} \approx - 7.88, \qquad
f_{KN\Sigma^*} \approx -3.22.
\end{equation}
In the numerical calculation, we use $f_{KN\Sigma^*} = -2.6$, which is
within the range of SU(3) symmetry breaking.

The diagrams of Figs.~\ref{fig:hyp}(c) and (e) can have $K$ interactions
with the baryon decuplet $\Sigma(1385)$ and $\Delta$.
To determine these couplings using SU(3) relations, we again start with the
$\pi\Delta\Delta$ interaction, which reads
\begin{equation}
\mathcal{L}_{\pi\Delta\Delta} = \frac{f_{\pi\Delta\Delta}}{M_\pi}
\overline{\Delta}^\alpha O_{\alpha\beta}(Z) \gamma_\mu \gamma_5
\bm{I}(\textstyle\frac32,\frac32) \cdot \partial^\mu \bm{\pi}
O^{\beta\delta}(Z) \Delta_{\delta}.
\end{equation}
By using the quark model prediction \cite{BW75}
$f_{\pi\Delta\Delta}/f_{\pi NN} = 4/5$ (
$f_{\pi NN} = g_{\pi NN} \frac{M_\pi}{2M_N}$), we find
$f_{\pi\Delta\Delta} \approx 0.8$.
The SU(3) relation
\begin{equation}
f_{K\Delta\Sigma^*} = -\sqrt{\frac32} f_{\pi\Delta\Delta}
\frac{M_K}{M_\pi} \approx 3.46,
\end{equation}
then fixes the following Lagrangian for $K\Delta\Sigma^*$ interaction 
\begin{equation}
\mathcal{L}_{K\Delta\Sigma^*} = \frac{f_{K\Delta\Sigma^*}}{M_K}
\overline{\Delta}^\alpha O_{\alpha\beta}(Z) \gamma_\mu \gamma_5
O^{\beta\delta}(Z)
\bm{I}(\textstyle\frac32,\frac12)
\cdot \bm{\Sigma}^*_\delta \partial^\mu K.
\end{equation}

Finally, we consider the interactions of baryon decuplet [$\Sigma^*(1358)$]
with the vector mesons ($K^*$) in Figs.~\ref{fig:hyp}(h) and (i).
In order to use the SU(3) symmetry relation, we start with the
$\rho N \Delta$ interaction,
\begin{equation}
\mathcal{L}_{\rho N\Delta} = i \frac{f_{\rho N\Delta}}{M_\rho}
\overline{\Delta}^\sigma O_{\sigma\mu}(Z) \gamma_5 \gamma_\nu
\bm{T}(\textstyle\frac32,\frac12) \cdot (\partial^\mu \bm{\rho}^\nu -
\partial^\nu \bm{\rho}^\mu ) N,
\end{equation}
where the coupling constant can be fixed by the quark model
relation~\cite{MHE87},
\begin{equation}
f_{\rho N\Delta} = \frac{f_{\pi N\Delta} g_{\rho NN}}{f_{\pi NN}}
\frac{M_\rho}{2M_N} (1+\kappa_\rho).
\end{equation}
By using $g_{\rho NN} = 3.1$ and $\kappa_\rho = 1.0$, we find
$f_{\rho N\Delta} \approx 5.5$.
This should be compared with the range $3.5$--$7.8$ of Ref.~\cite{KA04}.
By using the SU(3) relation
\begin{equation}
f_{K^*N\Sigma^*} = -\frac{f_{\rho N\Delta}}{\sqrt6}
\frac{M_{K^*}}{M_\rho} \approx 2.59,
\end{equation}
we then fix the $K^*N\Sigma^*$ coupling,
\begin{equation}
\mathcal{L}_{K^* N\Sigma^*} = i \frac{f_{K^*N\Sigma^*}}{M_{K^*}}
(\partial^\mu K^{*\nu} - \partial^\nu K^{*\mu})
\overline{\bm{\Sigma}}^{*\sigma} \cdot \bm{\tau}\, O_{\sigma\mu}
\gamma_5 \gamma_\nu N.
\end{equation}

With the above Lagrangians, we can evaluate all diagrams in
Fig.~\ref{fig:hyp} by specifying the propagator of spin-3/2
$\Sigma^*(1385)$ and $\Delta(1232)$.
Here we follow Refs.~\cite{ELP92,LLD91} and use the following
Rarita-Schwinger form for the propagator of a spin-3/2 particle of a mass
$M_{Y^*}$ and momentum $p$,
\begin{equation}
\tilde{\Delta}_{\mu\nu}(Y^*,p) = \frac{-i}{p^2-M_{Y^*}^2}
\Delta_{\mu\nu}(Y^*,p),
\end{equation}
where
\begin{equation}
\Delta_{\mu\nu}(Y^*,p) = (p\!\!\!/\, +
M_{Y^*}) S_{\mu\nu}(Y^*,p)
- \frac{2(p^2-M_{Y^*}^2)}{3M_{Y^*}^2}
\left[ \gamma_\mu p_\nu +
\gamma_\nu p_\mu - \gamma_\mu (p\!\!\!/\, - M_{Y^*}) \gamma_\nu \right],
\end{equation}
with
\begin{eqnarray}
S^{\mu\nu}(Y^*,p) &=& -g^{\mu\nu} + \frac13 \gamma^\mu \gamma^\nu +
\frac{1}{3M_{Y^*}} (\gamma^\mu p^\nu - \gamma^\nu p^\mu) +
\frac{2}{3M_{Y^*}^2} p^\mu p^\nu.
\end{eqnarray}
The decay width of the decuplet is included by replacing $M_{Y^*}^{}$ by
$M_{Y^*}^{} - i \Gamma_{Y^*}/2$ in the propagator.
Similar form of the propagator is also used for $N''=\Delta$ in evaluating
Fig.~\ref{fig:hyp}(c) and (e).

\subsection{$\bm{\Lambda(1405)}$}

The $\Lambda(1405)$ with $J^P = \frac12^-$ is also included in our
calculation of Fig.~\ref{fig:hyp}(a)-(e).
Its decay width is $\Gamma = 50 \pm 2$ MeV and it mostly decays into the
$\Sigma\pi$ channel.
To get its coupling with $K$, we consider scalar coupling SU(3)
Lagrangian,
\begin{equation}
\mathcal{L}_{\Lambda_1} = -ig_{\Lambda_1} \overline{\Lambda}_1 \left(
\overline{K} N + \overline{K}_c \Xi + \bm{\Sigma} \cdot \bm{\pi} +
\Lambda\eta \right) + \mbox{ H.c.},
\label{eq:LNK}
\end{equation}
where $\Lambda_1$ is the $\Lambda(1405)$ field.
This leads to
\begin{equation}
\Gamma\bm{(}\Lambda(1405) \to \Sigma\pi \bm{)} =
\frac{3g_{\Lambda_1}^2}{4\pi} \frac{p_\pi}{M_{\Lambda_1}} \left(
\sqrt{M_\Sigma^2 + p_\pi^2} + M_\Sigma \right),
\end{equation}
where $M_{\Lambda_1}$ is the mass of $\Lambda(1405)$ and $p_\pi =
\sqrt{\lambda(M_{\Lambda_1}^2,M_\Sigma^2,M_\pi^2)}/(2M_{\Lambda_1})$
with
\begin{equation}
\lambda(x,y,z) = x^2+y^2+z^2 - 2(xy+yz+zx).
\end{equation}
So using $\Gamma\bm{(}\Lambda(1405) \to \Sigma\pi \bm{)}_{\rm expt.}
\approx 50$ MeV, we obtain $ 
g_{\Lambda_1} \approx 0.13$ and the needed $KN\Lambda(1405)$
coupling can be computed from Eq.~(\ref{eq:LNK}).

Figure~\ref{fig:hyp}(d) can have $\gamma\Lambda_1\Lambda_1$ vertex.
This is calculated from 
\begin{equation}
\mathcal{L}_{\gamma\Lambda_1\Lambda_1} =
\frac{e\kappa_{\Lambda_1}}{2M_N} \overline{\Lambda}_1 \sigma_{\mu\nu}
\partial^\nu A^\mu \Lambda_1,
\end{equation}
where $\kappa_{\Lambda_1}$ is the magnetic moment of $\Lambda(1405)$.
There is no experimental information for the magnetic moment of
$\Lambda(1405)$. So we have to reply on model predictions.
It has been estimated to be $0.22 \sim 0.25$ in Skyrme model~\cite{SSG95}
and $0.24 \sim 0.45$ in a unitarized chiral perturbation theory~\cite{JHNO02}.
Here we use $\kappa_{\Lambda_1} = 0.25$ in the nucleon magneton unit.

For the intermediate hyperon state in Fig.~\ref{fig:hyp}(d), we also include
$\Lambda(1405) \to \Lambda(1116)\gamma$ and
$\Lambda(1405) \to \Sigma(1193)\gamma$ couplings.
They are defined by
\begin{eqnarray}
\mathcal{L}_{\gamma Y \Lambda_1} &=& \frac{eg_{\gamma Y
\Lambda_1}}{4(M_{\Lambda_1} + M_Y)} \overline{\Lambda}_1 \gamma_5
\sigma_{\mu\nu}Y F^{\nu\mu} + \mbox{ H.c.},
\end{eqnarray}
which leads to
\begin{equation}
\Gamma(\Lambda_1 \to Y\gamma) = \frac{\alpha_{\rm em} g_{\gamma
Y\Lambda_1}^2 p_\gamma^3}{(M_{\Lambda_1} + M_Y)^2},
\end{equation}
where $p_\gamma = (M_{\Lambda_1}^2 - M_Y^2)/(2M_{\Lambda_1})$.
By using the quark  model predictions
\cite{DHK83}
\begin{eqnarray}
&&
\Gamma\bm{(}\Lambda(1405) \to \Lambda\gamma \bm{)} = 143 \mbox{ keV},
\qquad
\Gamma\bm{(}\Lambda(1405) \to \Sigma^0\gamma \bm{)} = 91 \mbox{ keV},
\label{eq:qm1}
\end{eqnarray}
the parameters for $\mathcal{L}_{\gamma Y \Lambda_1}$ are then fixed as
\begin{equation}
g_{\gamma\Lambda\Lambda_1} \approx 2.67, \qquad
g_{\gamma\Sigma\Lambda_1} \approx 3.34.
\end{equation}
The predicted decay width $\Gamma\bm{(}\Lambda(1405) \to
\Sigma^{*0}(1385)\gamma \bm{)} = 0.3 \mbox{ keV}$ is suppressed and not
considered in this calculation.

\subsection{$\bm{\Lambda(1520)}$}

We now consider the calculations of Fig.~\ref{fig:hyp} with $Y=\Lambda(1520)$
 which is a $J^P = \frac32^-$ state. 
There is no information for the magnetic moment of $\Lambda(1520)$.
We therefore neglect the tensor coupling of the
electromagnetic interaction of $\Lambda(1520)$ by setting
$\mu[\Lambda(1520)]=0$.
The coupling of $\Lambda(1520)$ with photon thus has the same form as the
$\mathcal{L}_{\gamma\Sigma^*\Sigma^*}$ in Eq.~(\ref{eq:Lag-RS}) except
that $\Lambda(1520)$ is a neutral particle.

For the $KN\Lambda(1520)$ (denoted as $\Lambda^\prime$) and $\gamma K N
\Lambda(1520)$ couplings, we write
\begin{eqnarray}
\mathcal{L}_{KN\Lambda'} &=& \frac{f_{KN\Lambda'}}{M_K}
\overline{\Lambda}^\mu \gamma_5 \partial_\mu \overline{K} N +
\mbox{ H.c.},
\nonumber \\
\mathcal{L}_{\gamma KN\Lambda'} &=& ie \frac{f_{KN\Lambda'}}{M_K} A^\mu
\overline{\Lambda}_\mu \gamma_5 K^- p + \mbox{ H.c.}.
\end{eqnarray}
where $\Lambda^\mu$ is the $\Lambda(1520)$ field.
This gives the decay width of $\Lambda(1520) \to N\overline{K}$ as
\begin{equation}
\Gamma[\Lambda(1520) \to N \overline{K}] = \frac{1}{6\pi} \left(
\frac{f_{KN\Lambda'}}{M_K} \right)^2 \frac{p_K^3}{M_{\Lambda'}} \left(
\sqrt{M_N^2 + p_K^2} - M_N \right).
\end{equation}
Using $\Gamma[\Lambda(1520) \to N \overline{K}]_{\rm expt.} \approx 7$
MeV, we have
\begin{equation}
f_{KN\Lambda'} \approx 10.92.
\end{equation}

Figure~\ref{fig:hyp}(d) also includes the photon transition between
$\Lambda(1520)$ and other hyperons.
This is calculated by using the following Lagrangian~\cite{BMZ95},
\begin{eqnarray}
\mathcal{L}_{\gamma Y\Lambda'} = \frac{ie g_1}{2M_Y}
\overline{\Lambda}^\mu O_{\mu\lambda} \gamma_\nu Y F^{\nu\lambda}
 -\frac{e g_2}{4M_Y^2}
\overline{\Lambda}^\mu O_{\mu\nu} (\partial_\lambda Y) F^{\nu\lambda}.
\end{eqnarray}
Then we have
\begin{eqnarray}
\Gamma(\Lambda' \to Y\gamma) &=& \frac{\alpha_{\rm em} p_\gamma^3}{48
M_R^2 M_Y^2} \Biggl[ \left\{ g_1 (3M_R + M_Y) - g_2 \frac{M_R}{2M_Y} (M_R
- M_Y) \right\}^2
\nonumber \\ && \mbox{}
+ 3 \left( g_1 - g_2 \frac{M_R}{2M_Y} \right)^2 (M_R -
  M_Y)^2 - 8 g_2 M_R^2 \left( 2g_1 - g_2 \frac{M_R}{2M_Y} \right)
\Biggr].
\end{eqnarray}
For $\mathcal{L}_{\gamma \Lambda\Lambda'}$, we use the data
$\Gamma[\Lambda(1520) \to \Lambda\gamma]_{\rm expt.} = 159 \pm 33 \pm 26$
keV~\cite{SPHINX04a}%
\footnote{The new measurement of the CLAS collaboration~\cite{CLAS05a},
$167 \pm 43 \stackrel{+26}{\mbox{\scriptsize $-12$}}$ keV, is consistent
with the value of Ref.~\cite{SPHINX04a}.}
and set $g_2 = 0$ as in the case of $\Delta\to N\gamma$ to get
\begin{equation}
g_1 \approx 1.46.
\end{equation}
For the photo-transitions of $\Lambda(1520)$ to other hyperons, we again
are guided by the quark model predictions Ref.~\cite{DHK83},
\begin{eqnarray}
&&
\Gamma[\Lambda(1520) \to \Sigma\gamma] = 74 \mbox{ keV}, \qquad
\Gamma[\Lambda(1520) \to \Lambda(1405)\gamma] = 0.2 \mbox{ keV}, \qquad
\nonumber \\ &&
\Gamma[\Lambda(1520) \to \Sigma^*\gamma] \sim 0.
\end{eqnarray}
As the decay widths of $\Lambda(1520) \to \Sigma^*\gamma$ and
$\Lambda(1405)\gamma$ are negligible, we only
consider the coupling with the $\Sigma\gamma$
channel. Using $\Gamma[\Lambda(1520) \to \Sigma\gamma] = 74 \mbox{ keV}$
and also setting $g_2=0$, we get $\mathcal{L}_{\gamma \Sigma\Lambda'}$
with $ g_1 \approx 1.39$.

With the above four subsections, we have constructed the Lagrangians for
calculating all diagrams in Fig.~\ref{fig:hyp}.
We now turn to discussing $\Theta$ production mechanisms.

\section{Form factors and current conservation}

Here, we discuss how we restore current conservation with the form
factors in the form of Eq.~(\ref{eq:ff}).
Introducing form factors breaks current conservation.
So, motivated by the works of Refs.~\cite{Ohta89,Habe97,HBMF98a,DW01a},
we restore
current conservation by introducing contact diagrams, which
effectively changes the form factors of some diagrams into a universal form
factor.
For example, let us assume that we have two diagrams, $a$ and $b$, and,
when combined, these two diagrams satisfy current conservation without
form factors.
The corresponding form factors $F_a$ and $F_b$ as functions of
Mandelstam variables break current conservation.
Introducing contact diagrams effectively gives the universal form factor
which is a constant~\cite{Ohta89} or in the form of~\cite{HBMF98a},
\begin{equation}
F_a,F_b \to \frac12(F_a + F_b)
\end{equation}
so that the current conservation condition is satisfied.
It was pointed out by Ref.~\cite{DW01a} that such a choice would not
satisfy crossing symmetry and another form was suggested, although not
unique,
\begin{equation}
F_a,F_b \to 1 - (1 - F_a)(1 - F_b).
\end{equation}
In this work, since we have three-body final state, we extend the above
method to the diagrams which spoil current conservation due to form factors.
But we do not make any modification to the purely transverse amplitudes,
such as the terms with $K^*$ exchanges and photo-transitions among hadrons,
which are constructed to be gauge-invariant by themselves individually.
For later use, we define the Mandelstam variables as
\begin{eqnarray}
&& s = (k+p_1)^2, \quad s_1 = (q_1 + q_2)^2, \quad s_2 = (q_1 + p_2)^2,
\quad s_3 = (q_2 + p_2)^2, \nonumber \\
&& t_1 = (k-q_1)^2, \quad t_2 = (k-q_2)^2, \quad t_3 = (k-p_2)^2,
\nonumber
\\
&& t_4 = (p_1-q_1)^2, \quad t_5 = (p_1-q_2)^2, \quad t_6 = (p_1-p_2)^2,
\end{eqnarray}
where the momenta of the initial photon and the nucleon are $k$ and
$p_1$, respectively, while those of $\overline{K}$, $K$, and the final
nucleon are $q_1$, $q_2$, and $p_2$, respectively.

\subsection{$t$-channel Drell diagrams}

Among the possible (tree) $t$-channel Drell diagrams depicted in
Fig.~\ref{fig:Drell}, those diagrams with intermediate $K^*$ have
transverse amplitude only, i.e., each
diagram satisfied current conservation individually and introducing form
factors does not spoil the current conservation condition.
The current conservation problem occurs only when we have intermediate $K$
and $\overline{K}$ mesons, namely, $\gamma p \to K^+ K^- p$ and $\gamma
n \to K^+ K^- n$, since they do not contribute to the reactions of
$\gamma p \to K^0 \overline{K}^0 p$ and
$\gamma n \to K^0 \overline{K}^0 n$.
Then the amplitude takes a form of
\begin{equation}
\mathcal{M}^\mu =
\mathcal{M}_a^\mu F_a + \mathcal{M}_b^\mu F_b + \mathcal{M}_c^\mu F_c,
\label{eq:Dr}
\end{equation}
where
\begin{equation}
F_a = \mathcal{F}(t_2,M_K^2) \mathcal{F}(t_6,M_V^2), \quad
F_b = \mathcal{F}(t_1,M_K^2) \mathcal{F}(t_6,M_V^2), \quad
F_c = \mathcal{F}(t_6,M_V^2),
\end{equation}
where
\begin{equation}
\mathcal{F}(t_2,M_K^2) = F(t_2,M_K^2)^2, \quad
\mathcal{F}(t_1,M_K^2) = F(t_1,M_K^2)^2, \quad
\mathcal{F}(t_6,M_V^2) = F(t_6,M_V^2)^2,
\end{equation}
and the form of $F(r,M^2)$ is defined in Eq.~(\ref{eq:ff}).
One can verify that the amplitude (\ref{eq:Dr}) satisfies current
conservation condition when the form factors are set to be 1,
\begin{equation}
k \cdot \mathcal{M} = k \cdot ( \mathcal{M}_a +
\mathcal{M}_b + \mathcal{M}_c) = 0.
\end{equation}
The condition, $k \cdot \mathcal{M}$, is satisfied by introducing the contact
diagrams, which effectively replaces $F_a$, $F_b$, and $F_c$ by
\begin{equation}
1 - (1-F_a)(1-F_b)(1-F_c).
\end{equation}

\subsection{Vector meson and Tensor meson parts}

Since the form factors and their cutoff parameters are fixed by vector
meson photoproduction and tensor meson photoproduction processes where
the vector or tensor mesons are on mass-shell, we need to include the
form factors in order to take into account the off-shell--ness of the
intermediate vector/tensor mesons.
Furthermore, the amplitudes $\mathcal{M}^{\mu\nu}$ of
Eq.~(\ref{eq:vm-part}) and $\mathcal{M}^{\mu,\alpha\beta}$ of
Eq.~(\ref{eq:tm-part}) are constructed to satisfy the current
conservation condition.
So the form factor which should be multiplied in addition to the form
factors for vector/tensor meson photoproduction part reads
\begin{equation}
F = F(s_1, M^2)^2,
\end{equation}
where $M$ is the (produced) vector meson or tensor meson mass.

\subsection{Intermediate hyperons}

\subsubsection{Intermediate spin-1/2 hyperons}

In this case, since we adopt the pseudoscalar coupling, we have five diagrams
as shown in Figs.~\ref{fig:hyp}(a)-(e).
The possible intermediate states are $\Lambda(1116)$, $\Lambda(1405)$,
and $\Sigma(1193)$.
We first consider the case of $\Sigma(1193)$, which have the properties as
\begin{eqnarray}
&& k \cdot \mathcal{M}_c^n = 0, \qquad
 k \cdot \mathcal{M}_d^{\Sigma^0} = 0, \qquad
 k \cdot \mathcal{M}_e^n = 0,
\nonumber \\ &&
k \cdot \mathcal{M}_d^{\Sigma^+} =
k\cdot \mathcal{M}_b + k \cdot \mathcal{M}_a =
-k\cdot \mathcal{M}_e^p - k \cdot \mathcal{M}_c^p,
\nonumber \\ &&
k \cdot \mathcal{M}_a + k \cdot \mathcal{M}_c^p = 0, \qquad
 k \cdot \mathcal{M}_b + k \cdot \mathcal{M}_e^p = 0,
\label{eq:pr1}
\end{eqnarray}
where the superscripts $p,n$ denote the proton and neutron, respectively,
and the subscripts specify the diagram in Fig.~\ref{fig:hyp}.

For $\gamma p \to K^+ K^- p$, we have
\begin{equation}
\mathcal{M}^\mu = \mathcal{M}_a^\mu F_a + \mathcal{M}_b^\mu F_b +
\mathcal{M}_c^{p,\mu} F_c + \mathcal{M}_d^{\Sigma^0,\mu} F_d
+ \mathcal{M}_e^{p,\mu} F_e.
\label{eq:Sig1}
\end{equation}
Because of the properties of Eq.~(\ref{eq:pr1}), we have $k \cdot
\mathcal{M} = 0$ without form factors.
Since the amplitudes $\mathcal{M}_a$ and $\mathcal{M}_c$ correspond to
a part of $\Lambda$ photoproduction, which should have nontrivial form factors,
we replace the form factors as
\begin{equation}
F_a, F_c \to \{ 1 - (1 - F(t_2,M_K^2)^2)(1 - F(s,M_N^2)^2) \}
F(s_2,M_\Sigma^2)^2.
\end{equation}
Similarly,
\begin{equation}
F_b, F_e \to \{ 1 - (1 - F(t_1,M_K^2)^2)(1 - F(t_3,M_N^2)^2) \}
F(t_5,M_\Sigma^2)^2,
\end{equation}
so that current conservation is now satisfied with the form factors.

For $\gamma p \to K^0 \overline{K}^0 p$, we have
\begin{equation}
\mathcal{M}^\mu = 2 (\mathcal{M}_c^{p,\mu} F_c +
\mathcal{M}_d^{\Sigma^+,\mu} F_d + \mathcal{M}_e^{p,\mu} F_e),
\end{equation}
where the form factors $F_c$, $F_d$, and $F_e$ are replaced by
\begin{equation}
1 - (1-F_c)(1-F_d)(1-F_e).
\end{equation}

For $\gamma n \to K^+ K^- n$, we have
\begin{equation}
\mathcal{M}^\mu = 2 (\mathcal{M}_a^\mu F_a + \mathcal{M}_b^\mu F_b +
\mathcal{M}_c^{n,\mu} F_c + \mathcal{M}_d^{\Sigma^-,\mu} F_d
+ \mathcal{M}_e^{n,\mu} F_e),
\label{eq:Sig3}
\end{equation}
where the form factors $F_a$, $F_b$, and $F_d$ are replaced by
\begin{equation}
1 - (1-F_a)(1-F_b)(1-F_d).
\end{equation}
Since $\mathcal{M}_c^{n,\mu}$ and $\mathcal{M}_e^{n,\mu}$ are
transverse, the form factors $F_c$ and $F_e$ are not changed.

For $\gamma n \to K^0 \overline{K}^0 n$, we have
\begin{equation}
\mathcal{M}^\mu = 
\mathcal{M}_c^{n,\mu} F_c + \mathcal{M}_d^{\Sigma^0,\mu}  F_d
+ \mathcal{M}_e^{n,\mu} F_e.
\end{equation}
Since all the amplitudes are transverse, they are gauge invariant.

The intermediate $\Lambda(1116)$ and $\Lambda(1405)$ states are the same
as the case of $\Sigma^0$.
As a consequence, these intermediate states do not exist for the
$\gamma p \to K^0 \overline{K}^0 p$ and $\gamma n \to K^+ K^- n$ reactions.
We replace the form factors as in Eq.~(\ref{eq:Sig1}) for $\gamma p \to
K^+ K^- p$. The amplitudes for $\gamma n \to K^0 \overline{K}^0 n$ are
all transverse.

\subsubsection{Intermediate spin-3/2 hyperons}

The intermediate states $\Sigma(1385)$ and $\Lambda(1520)$
contain seven diagrams shown in
Figs.~\ref{fig:hyp}(a)-(g), which satisfy
\begin{eqnarray}
&& k \cdot \mathcal{M}_c^n = 0, \qquad
k \cdot \mathcal{M}_d^{\Sigma^{*0}} = 0, \qquad
k \cdot \mathcal{M}_e^n = 0,
\nonumber \\ &&
k \cdot \mathcal{M}_d^{\Sigma^{*+}} =
-k \cdot \mathcal{M}_d^{\Sigma^{*-}} = - k \cdot \mathcal{M}_c^p - k
\cdot \mathcal{M}_e^p,
\nonumber \\ &&
k \cdot \mathcal{M}_a + k \cdot \mathcal{M}_c^p + k \cdot \mathcal{M}_f
= 0,
\nonumber \\ &&
k \cdot \mathcal{M}_b + k \cdot \mathcal{M}_e^p + k \cdot \mathcal{M}_g
= 0,
\end{eqnarray}
for the case of intermediate $\Sigma^*(1385)$.

For $\gamma p \to K^+ K^- p$, we have
\begin{equation}
\mathcal{M}^\mu = \mathcal{M}_a^\mu F_a + \mathcal{M}_b^\mu F_b +
\mathcal{M}_c^{p,\mu} F_c + \mathcal{M}_d^{\Sigma^{*0},\mu} F_d +
\mathcal{M}_e^{p,\mu} F_e + \mathcal{M}_f^\mu F_f + \mathcal{M}_g^\mu
F_g.
\end{equation}
In this case, the direct application of the method of Ref.~\cite{DW01a},
\begin{equation}
F_a, F_c, F_f \to 1 - (1-F_a)(1-F_c)(1-F_f), \qquad
F_b, F_e, F_g \to 1 - (1-F_b)(1-F_e)(1-F_g),
\label{eq:try1}
\end{equation}
cannot be used.
This is because the above form factor becomes 1 if, for example, any of
($F_a$, $F_c$, $F_f$) is 1.
And this in fact happens when the intermediate $\Sigma^*$ becomes on
mass-shell, $s_2 = M_{\Sigma^*}^2$ since $F_f = F(s_2,M_{\Sigma^*}^2)^2$.
Since these diagrams are a part of photoproduction of $K$ and
$\Sigma^*$, this means that those amplitudes have no form factor.
Therefore, instead of Eq.~(\ref{eq:try1}), we replace the form factors as
\begin{eqnarray}
F_a, F_c, F_f &\to& \{ 1 - (1 - F(t_2,M_K^2)^2)(1 - F(s,M_N^2)^2) \}
F(s_2,M_{\Sigma^*}^2)^2,
\nonumber \\
F_b, F_e, F_g &\to& \{ 1 - (1 - F(t_1,M_K^2)^2)(1 - F(t_3,M_N^2)^2) \}
F(t_5,M_{\Sigma^*}^2)^2.
\end{eqnarray}

For $\gamma p \to K^0 \overline{K}^0 p$, we have
\begin{equation}
\mathcal{M}^\mu = 2 (\mathcal{M}_c^{p,\mu} F_c +
\mathcal{M}_d^{\Sigma^{*+},\mu} F_d + \mathcal{M}_e^{p,\mu} F_e).
\end{equation}
The form factors $F_c$, $F_d$, and $F_e$ are replaced by
\begin{equation}
1 - (1-F_c)(1-F_d)(1-F_e).
\end{equation}

For $\gamma n \to K^+ K^- n$, we have
\begin{equation}
\mathcal{M}^\mu = 2 (\mathcal{M}_a^\mu F_a + \mathcal{M}_b^\mu F_b +
\mathcal{M}_c^{n,\mu} F_c + \mathcal{M}_d^{\Sigma^{*-},\mu} F_d +
\mathcal{M}_e^{n,\mu} F_e + \mathcal{M}_f^\mu F_f + \mathcal{M}_g^\mu F_g ),
\end{equation}
where the form factors $F_a$, $F_b$, $F_d$, $F_f$, and $F_g$ are
\begin{equation}
1 - (1-F_a)(1-F_b)(1-F_d)(1-F_f')(1-F_g'),
\end{equation}
and
\begin{equation}
F_f' = F_g' = F_f F_g,
\end{equation}
which is chosen to avoid the problem mentioned above.

For $\gamma n \to K^0 \overline{K}^0 n$, we have
\begin{equation}
\mathcal{M}^\mu = 
\mathcal{M}_c^{n,\mu} F_c + \mathcal{M}_d^{\Sigma^{*0},\mu}  F_d
+ \mathcal{M}_e^{n,\mu} F_e.
\end{equation}
In this case, all amplitudes are gauge-invariant, we do not introduce
any contact diagrams.

The case of intermediate $\Lambda(1520)$ is the same as in the case of
$\Sigma^0(1385)$.
So it does not contribute to the $\gamma p \to K^0 \overline{K}^0 p$ and
$\gamma n \to K^+ K^- n$ reactions.

\subsection{Intermediate $\Theta$}

This case is similar to the case of intermediate $\Lambda$ except that
$\Theta$ has positive charge.
The $K^*$ exchange diagrams, Figs.~\ref{fig:theta}(f,g), are transverse,
so current conservation is satisfied even with the presence of form factors.
The amplitudes of Fig.~\ref{fig:theta} satisfy
\begin{eqnarray}
&& k \cdot \mathcal{M}_c^n = 0, \qquad
k \cdot \mathcal{M}_e^n = 0,
\nonumber \\ &&
k \cdot \mathcal{M}_a + k \cdot \mathcal{M}_b + k \cdot \mathcal{M}_d =
0,
\nonumber \\ &&
k \cdot \mathcal{M}_c + k \cdot \mathcal{M}_d + k \cdot \mathcal{M}_e =
0.
\end{eqnarray}

For $\gamma p \to K^0 \overline{K}^0 p$, we have
\begin{equation}
\mathcal{M}^\mu = 
\mathcal{M}_c^{p,\mu} F_c + \mathcal{M}_d^\mu F_d +
\mathcal{M}_e^{p,\mu} F_e + \mathcal{M}_f^\mu F_f + \mathcal{M}_g^\mu
F_g,
\end{equation}
where
\begin{equation}
F_c, F_d, F_e \to 1 - (1-F_c)(1-F_d)(1-F_e).
\end{equation}

For $\gamma n \to K^+ K^- n$, we have
\begin{equation}
\mathcal{M}^\mu =  \mathcal{M}_a F_a + \mathcal{M}_b F_b
\mathcal{M}_c^{n,\mu} F_c + \mathcal{M}_d^\mu F_d +
\mathcal{M}_e^{n,\mu} F_e + \mathcal{M}_f^\mu F_f + \mathcal{M}_g^\mu
F_g,
\end{equation}
where
\begin{equation}
F_a, F_b, F_d \to 1 - (1-F_a)(1-F_b)(1-F_d).
\end{equation}

%

\end{document}